\input harvmac
\input epsf
\input tables
\newif\ifdraft\draftfalse
\newif\ifinter\interfalse
\ifdraft\draftmode\else\interfalse\fi
\def\journal#1&#2(#3){\unskip, \sl #1\ \bf #2 \rm(19#3) }
\def\andjournal#1&#2(#3){\sl #1~\bf #2 \rm (19#3) }

\def\frac#1#2{{#1\over#2}}

\def\d{\partial}

\def\inbar{\,\vrule height1.5ex width.4pt depth0pt}
\def\IC{\relax\hbox{$\inbar\kern-.3em{\rm C}$}}
\def\IR{\relax{\rm I\kern-.18em R}}
\def\IP{\relax{\rm I\kern-.18em P}}

%
%


%
\catcode`\@=11
\def\slash#1{\mathord{\mathpalette\c@ncel{#1}}}
\overfullrule=0pt

\def\DD{{\cal D}}

\def\FF{{\cal F}}

\def\HH{{\cal H}}

\def\MM{{\cal M}}
\def\NN{{\cal N}}
\def\OO{{\cal O}}

\def\SS{{\cal S}}

\def\UU{{\cal U}}

\def\underrel#1\over#2{\mathrel{\mathop{\kern\z@#1}\limits_{#2}}}

\catcode`\@=12


%

\def\tr{{\rm Tr}}

\def \sinh{{\rm sinh}}
\def \cosh{{\rm cosh}}


\def\opo{{1+1}}

\def\tbar{{\bar\theta}}

\def\[{[}
\def\]{]}

\def\comment#1{ }

%
\def\draftnote#1{\ifdraft{\baselineskip2ex
                 \vbox{\kern1em\hrule\hbox{\vrule\kern1em\vbox{\kern1ex
                 \noindent \underbar{NOTE}: #1
             \vskip1ex}\kern1em\vrule}\hrule}}\fi}
\def\internote#1{\ifinter{\baselineskip2ex
                 \vbox{\kern1em\hrule\hbox{\vrule\kern1em\vbox{\kern1ex
                 \noindent \underbar{Internal Note}: #1
             \vskip1ex}\kern1em\vrule}\hrule}}\fi}

%
%






\def\om{\omega}               
               

%
%
\def\inbar{\hskip.3em\vrule height1.5ex width.4pt depth0pt}
\def\IC{\relax{\inbar\kern-.3em{\rm C}}}
\def\IN{\relax{\rm I\kern-.16em N}}
\def\IQ{\relax\hbox{$\inbar$\kern-.3em{\rm Q}}}
\def\IZ{\relax{\rm Z\kern-.8em Z}}
%
%

%


\def\boxy{\epsfbox{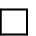}}

%
\def\bUU{\bar {{\cal U}}}

\def\bO{\bar{\Omega}}
\def\bD{\bar D}
\def\D{\Delta}
\def\bPsi{\bar {\Psi}}
\def\tp{\widehat{+}}
\def\tm{\widehat{-}}
\rightline{EFI-02-47}
\rightline{HU-EP-02/55}
\Title{
\rightline{hep-th/0212111}}
{\vbox{\centerline{BMN Operators for ${\cal N}=1$ Superconformal}
\vskip 10pt
\centerline{Yang-Mills Theories and Associated}
\vskip 10pt
\centerline{String Backgrounds}}}
\bigskip
\centerline{Vasilis Niarchos $^a$ and Nikolaos Prezas $^ b$}
\bigskip
\centerline{\it $^a$ Department of Physics, University of Chicago}
\centerline{\it 5640 S. Ellis Av., Chicago, IL 60637, USA}
\centerline{vniarcho@theory.uchicago.edu}
\centerline{\it $^b$ Institut f\"ur Physik, Humboldt Universit\"at zu Berlin}
\centerline{\it Invalidenstra\ss e 110, 10115 Berlin, Germany}
\centerline{ prezas@physik.hu-berlin.de}

\bigskip
\noindent

We study a class of near-BPS operators for a complex 2-parameter family of ${\cal N}=1$ superconformal Yang-Mills
theories that can be obtained by a Leigh-Strassler deformation of ${\cal N}=4$ SYM theory.
We identify these operators in the large $N$ and large $R$-charge limit and compute their exact scaling dimensions
using ${\cal N}=1$ superspace methods. From these scaling dimensions we attempt to reverse-engineer the light-cone
worldsheet theory that describes string propagation on the Penrose limit of the dual geometry.

\vfill

\Date{December, 2002}


\lref\bmn{D. Berenstein, J. Maldacena, H. Nastase, ``Strings in flat space and pp waves from $\NN =4$ Super
Yang Mills'', hep-th/0202021.}

\lref\gomis{J. Gomis, H. Ooguri, ``Penrose limit of $\NN =1$ gauge theories'', hep-th/0202157.}

\lref\mukhi{N. Itzhaki, I. R. Klebanov, S. Mukhi, ``PP wave limit and enhanced supersymmetry in gauge theories'',
hep-th/0202153.}

\lref\zayas{L. A. Pando Zayas, J. Sonnenschein, ``On Penrose limits and gauge
theories'', hep-th/0202186.}

\lref\onepapado{M. Blau, J. Figueroa-O'Farill, C. Hull, G. Papadopoulos, ``A new maximally supersymmetric background
of IIB superstring theory'', JHEP 0201 (2002) 047, hep-th/0110242.}

\lref\twopapado{M. Blau, J. Figueroa-O'Farill, C. Hull, G. Papadopoulos, ``Penrose limits and maximal supersymmetry'',
hep-th/0201081.}

\lref\threepapado{M. Blau, J. Figueroa-O'Farill, G. Papadopoulos, ``Penrose limits, supergravity and brane dynamics'',
hep-th/0202111.}

\lref\metsaev{R. R. Metsaev, ``Type IIB Green-Schwarz superstring in plane wave Ramond-Ramond background'',
Nucl. Phys. B 625 (2002) 70, hep-th/0112044.}

\lref\mettseyt{R. R. Metsaev, A. A. Tseytlin, ``Exactly solvable model of superstring in plane wave Ramond-Ramond
background'', hep-th/0202109.}

\lref\malda{J. Maldacena, ``The large N limit of superconformal field theories and supergravity'', Adv. Theor.
Math. Phys. 2 (1998) 231, hep-th/9711200.}

\lref\gubser{S. S. Gubser, I. R. Klebanov, A. M. Polyakov, ``Gauge theory correlators from non-critical string theory,''
Phys. Lett. B428 (1998) 105, hep-th/9802109.}

\lref\adswitten{E. Witten, ``Anti-de Sitter space and holography'', Adv. Theor. Math. Phys. 2 (1998) 253, hep-th/9802150.}

\lref\penrose{R. Penrose, ``Any space-time has a plane wave as a limit'', in Differential Geometry and relativity, 271-275,
Reidel, Dordrecht, 1976.}

\lref\leigh{R. G. Leigh, M. J. Strassler, ``Exactly marginal operators and duality in four-dimensional $\NN=1$
supersymmetric gauge theory,'' Nucl. Phys. B447 (1995) 95, hep-th/9503121.}

\lref\bigads{O. Aharony, S. S. Gubser, J. M. Maldacena, H. Ooguri, Y. Oz, ``Large $N$ field theories,
string theory and gravity'', Phys. Rept. 323 (2000) 183, hep-th/9905111.}

\lref\novel{L. Girardello, M. Petrini, M. Porrati, A. Zaffaroni, ``Novel local CFT and exact results on
perturbations of $\NN=4$ super Yang-Mills from $AdS$ dynamics'', JHEP 12 (1998) 022, hep-th/9810126.}

\lref\fayya{A. Fayyazuddin, S. Mukhopadhyay, ``Marginal perturbations of $\NN=4$ Yang-Mills as deformations
of $AdS_5 \times S^5$'', hep-th/0204056.}

\lref\polstrass{J. Polchinski, M. J. Strassler, ``The string dual of a confining four-dimensional gauge theory'',
hep-th/0003136.}

\lref\aharony{O. Aharony, B. Kol, S. Yankielowicz, ``On exactly marginal deformations of $\NN=4$ SYM and
type IIB supergravity on $AdS_5 \times S^5$'', hep-th/0205090.}

\lref\karch{A. Karch, D. Lust, A. Miemiec, `` New $\NN=1$ superconformal field theories and their
supergravity description'', hep-th/9901041.}

\lref\warner{A. Khavaev, K. Pilch, N. P. Warner, ``New vacua of gauged $\NN=8$ supergravity in five dimensions'',
hep-th/9812035.}

\lref\freedman{D. Z. Freedman, S. S. Gubser, K. Pilch, N. P. Warner, ``Renormalization group flows from holography-
supersymmetry and a c-theorem'', hep-th/9904017.}

\lref\ahakachru{O. Aharony, S. Kachru, E. Silverstein, ``unpublished''. Some of the results were presented by O. Aharony
at the ITP program on dualities in string theory, April 1998 and at Strings `98, June 1998, http://www.itp.ucsb.edu/
online/strings98/aharony and appeared in \bigads.}

\lref\distler{J. Distler, F. Zamora, ``Non-supersymmetric conformal field theories from stable Anti-de Sitter spaces'',
Adv. Theor. Math. Phys. 2 (1998) 1405, hep-th/9810206.}

\lref\nsvz{V. Novikov, M. A. Shifman, A. I. Vainshtein, V. Zakharov, ``Exact Gell-Mann-Low function of supersymmetric
Yang-Mills theories from instanton calculus'', Nucl. Phys. B229 (1983) 381.}

\lref\romans{H. J. Kim, L. J. Romans, P. van Nieuwenhuizen, ``The mass spectrum of chiral $\NN=2$ $d=10$
supergravity on $S^5$'', Phys. Rev. D32 (1985) 389.}

\lref\grana{M. Grana, J. Polchinski, ``Supersymmetric three-form flux perturbations on $AdS_5$'', Phys. Rev. D63
(2001) 026001, hep-th/0009211.}

\lref\intriligator{K. A. Intriligator, ``Bonus symmetries of $\NN=4$ super-Yang-Mills correlation functions via
$AdS$ duality'', Nucl. Phys. B551 (1999) 575, hep-th/9811047.}

\lref\onemaldasl{J. Maldacena, H. Ooguri, ``Strings in $AdS_3$ and $SL(2,R)$ WZW model 1: the
spectrum'', J. Math. Phys. 42: 2929-2960 (2001), hep-th/0001053.}

\lref\twomaldasl{J. Maldacena, H. Ooguri, J. Son, ``Strings in $AdS_3$ and $SL(2,R)$ WZW model. Part 2:
Euclidean Black Hole'', J. Math. Phys. 42: 2961-2977 (2001), hep-th/0005183.}

\lref\threemaldasl{J. Maldacena, H. Ooguri, ``Strings in $AdS_3$ and $SL(2,R)$ WZW model. Part 3: Correlation
Functions'', hep-th/0111180.}

\lref\kutrasone{A. Giveon, D. Kutasov, N. Seiberg, ``Comments on string theory on $AdS_3$'', Adv. Theor. Math. Phys. 2 (1998)
733, hep-th/9806194.}

\lref\kutrastwo{D. Kutasov, N. Seiberg, ``More comments on string theory on $AdS_3$'', JHEP 9904 (1999) 008, hep-th/9903219.}

\lref\kutrasthree{A. Giveon, D. Kutasov, ``Notes on $AdS_3$'', hep-th 0106004.}

\lref\alisha{M. Alishahiha, M. M. Sheikh-Jabbari, ``The PP-wave limits of orbifolded $AdS_5 \times S^5$'', hep-th/0203018.}

\lref\rey{N. W. Kim, A. Pankiewicz, S. J. Rey, S. Theisen, ``Superstring on pp-wave orbifold from large-N quiver
gauge theory'', hep-th/0203080.}

\lref\taka{T. Takayanagi, S. Terashima, ``Strings on orbifolded pp-waves'', hep-th/0203093.}

\lref\kehagias{E. Floratos, A. Kehagias, ``Penrose limits of orbifolds and orientifolds'', hep-th/0203134.}

\lref\chu{C. S. Chu, P. M. Ho, ``Noncommutative D-brane and open string in pp-wave background with B-field'', hep-th/0203186.}

\lref\gava{D. Berenstein, E. Gava, J. Maldacena, K. S. Narain, H. Nastase, ``Open strings on plane waves and their
Yang-Mills duals'', hep-th/0203249.}

\lref\bala{V. Balasubramanian, M. Huang, T. S. Levi, A. Naqvi, ``Open strings from $\NN =4$ SYM'', hep-th/0204196.}

\lref\andrei{A. Parnachev, D. Sahakyan, ``Penrose limit and string quantization in $AdS_3 \times S^3$, hep-th/0205015.}

\lref\andryzhov{A. Parnachev, A. Ryzhov, ``Strings in the near plane wave background and AdS/CFT'', hep-th/0208010.}

\lref\staudacher{C. Kristjansen, J. Plefka, G. W. Semenoff, M. Staudacher, ``A new double-scaling limit of
$\NN=4$ super Yang-Mills theory and pp-wave strings'', hep-th/0205033.}

\lref\gross{D. J. Gross, A. Mikhailov, R. Roiban, ``Operators with large $R$ charge in $\NN=4$ Yang-Mills
theory'', hep-th/0205066.}

\lref\motl{N. R. Constable, D. Z. Freedman, M. Headrick, S. Minwalla, L. Motl, A. Postnikov, W. Skiba,
``pp-wave string interactions from perturbative Yang-Mills theory'', hep-th/0205089.}

\lref\pilch{K. Pilch, N. P. Warner, ``A new supersymmetric compactification of chiral IIB supergravity'',
Phys. Lett. B 487, 22 (2000), hep-th/0002192.}

\lref\corrado{R. Corrado, N. Halmagyi, K. Kennaway, N. P. Warner, ``Penrose limits of RG fixed points and
pp-waves with background fluxes'', hep-th/0205314.}

\lref\gimon{E. Gimon, L. A. Pando Zayas, J. Sonnenschein, ``Penrose limits and RG flows'', hep-th/0206033.}

\lref\niarchos{V. Niarchos, work in progress.}

\lref\verlinde{H. Verlinde, ``Bits, Matrices and 1/N'', hep-th/0206059.}

\lref\gueven{R. Gueven, ``Plane wave limits and T-duality'', hep-th/0005061.}

\lref\myers{D. Brecher, C. V. Johnson, K. J. Lovis, R. C. Myers, ``Penrose limits, deformed pp-waves and
the string duals of $\NN=1$ large $N$ gauge theory'', hep-th/0206045.}

\lref\cvetic{M. Cvetic, H. L\''u, C. N. Pope, ``M-theory pp-waves, Penrose limits and supernumerary supersymmetries'',
hep-th/0203229.}

\lref\italian{A. Santambrogio, D. Zanon, ``Exact anomalous dimensions of ${\cal N}=4$ Yang-Mills operators with
large $R$ charge'', hep-th/0206079.}

\lref\tseytlin{J. G. Russo, A. A. Tseytlin, ``On solvable models of type IIB superstring in
NS-NS and R-R plane wave backgrounds'', JHEP 0204:021 (2002), hep-th/0202179.}

\lref\kol{B. Kol, ``On conformal deformations'', hep-th/0205141.}

\lref\santapenati{S. Penati, A. Santambrogio, ``Superspace approach to anomalous dimensions in $\NN=4$ SYM'',
hep-th/0107071.}

\lref\petkou{G. Arutyunov, S. Penati, A. C. Petkou, A. Santambrogio, E. Sokatchev, ``Non-protected operators in
$\NN=4$ SYM and multiparticle states of AdS$_5$ SUGRA'', hep-th/0206020.}

\lref\dalaras{A. Ceresole, G. Dall'Agata, R. D'Auria, S. Ferrara, ``Spectrum of type IIB supergravity on
$AdS_5 \times S^5$: Predictions on $\NN=1$ SCFT's'', hep-th/9905226.}

\lref\wb{J. Bagger, J. Wess, ``Supersymmetry and Supergravity'', Princeton series in physics,
Princeton (1992).}

\lref\schwarz{J. H. Schwarz, ``Covariant field equations of chiral $\NN=2$ $D=10$ supergravity'',
Nucl. Phys. B 226:269 (1983).}

\lref\fourpapado{J. Figueroa-O'Farill, G. Papadopoulos,``
Homogeneous fluxes, branes and a maximally supersymmetric solution of
M-theory'', JHEP 0108:036 (2001), hep-th/0105308}

\lref\beisert{N. Beisert, ``BMN operators and superconformal symmetry'', hep-th/0211032.}

\lref\gursoy{U. Gursoy, ``Vector operators in the BMN correspondence'', hep-th/0208041.}

\lref\staudachertwo{N. Beisert, C. Kristjansen, J. Plefka, G. W. Semenoff, M. Staudacher, ``BMN correlators
and operator mixing in $\NN=4$ super Yang-Mills theory'', hep-th/0208178.}

\lref\headrick{N. R. Constable, D. Z. Freedman, M. Headrick, S. Minwalla, ``Operator mixing and the BMN
correspondence'', hep-th/0209002.}

\lref\roiban{D. J. Gross, A. Mikhailov, R. Roiban, ``A calculation of the plane wave string Hamiltonian from
$\NN=4$ super-Yang-Mills theory'', hep-th/0208231.}

\lref\volovich{J. Pearson, M. Spradlin, D. Vaman, H. Verlinde, A. Volovich, ``Tracing the string: BMN
correspondence at finite $J^2/N$'', hep-th/0210102.}

\lref\razamat{O. Aharony, S. S. Razamat, ``Exactly marginal deformations of $\NN=4$ SYM
and of its supersymmetric orbifold descendants'', JHEP 0205:029 (2002), hep-th/0204045.}

\lref\osborn{H. Osborn, ``$\NN=1$ superconformal symmetry in four-dimensional quantum field theory'',
Annals Phys. 272 (1999) 243, hep-th/9808041.}

\lref\osborntwo{F. A. Dolan, H. Osborn, ``On short and semi-short
representations for four-dimensional superconformal symmetry'', hep-th/0209056.}

\lref\razamattwo{S. S. Razamat, ``Marginal deformations of $\NN=4$ SYM and of its supersymmetric
orbifold descendants'', hep-th/0204043.}

\lref\beresone{D. Berenstein, V. Jejjala, R. G. Leigh, ``Marginal and relevant deformations of
$\NN=4$ field theories and non-commutative moduli spaces of vacua'', hep-th/0005087.}

\lref\berestwo{D. Berenstein, R. G. Leigh, ``Discrete torsion, AdS/CFT and duality'', hep-th/0001055.}

\lref\oz{Y. Oz, T. Sakai, ``Exact anomalous dimensions for $\NN=2$ ADE SCFTs'',
hep-th/0208078.}

\lref\IntriligatorJJ{
K.~Intriligator and B.~Wecht,
``The exact superconformal R-symmetry maximizes a,''
arXiv:hep-th/0304128.
}

\newsec{Introduction}

According to the AdS/CFT correspondence \refs{\malda,\gubser,\adswitten} string theory
on spaces of the form $AdS_d \times \MM^{10-d}$ is dual to a
conformal field theory that lives on the $(d-1)$-dimensional boundary of $AdS_d$.
Several examples of this correspondence have been studied so far.
From the CFT point of view
new conformal or nonconformal examples can be obtained by
deforming the gauge theory action with a local operator $\OO$
\eqn\deformgeneral{S \rightarrow S+h\int d^d x \OO(x).}
Usually such deformations break some or all of the initial supersymmetry
and in most cases it is a nontrivial task to determine how this deformation reflects
itself on the string theory side.
When $\OO$ is a relevant operator
the deformation breaks conformal invariance and the RG flow can lead to an
interacting IR fixed point. On the gravity side
such a deformation yields a complicated space with a running dilaton that interpolates between two $AdS$ geometries of the
form $AdS \times \MM_{UV}$ and $AdS \times \MM_{IR}$.
When $\OO$ is exactly marginal, conformal invariance
remains as a true symmetry of the theory and the dual geometry takes the form $AdS \times \MM_h$, with
$\MM_h$ a compact deformed version of the original manifold $\MM$.

We are interested in type IIB string theory on $AdS_5 \times S^5$ and
deformations of its dual $\NN=4$ SYM.  Many interesting papers have been written on this subject.
For example, non-supersymmetric deformations were discussed in \distler. Exactly marginal and relevant
deformations that preserve $\NN=1$ supersymmetry
were discussed in \refs{\ahakachru\novel\karch\freedman\berestwo\beresone\fayya-\aharony}
(for a brief review see
\bigads, section 4.3).
A relevant perturbation that leads to a confining gauge theory was discussed in \polstrass.
All these cases were considered in the large t'Hooft limit, where supergravity is reliable.

Here we want to discuss a certain class of $\NN=1$ superconformal Yang-Mills theories that can be
obtained by a Leigh-Strassler deformation of the $\NN=4$ SYM theory \refs{\leigh,\aharony}. Our analysis
focuses on the properties of various near-BPS gauge theory operators with large $R$-charge. These operators were
considered recently by the authors of \bmn, who also proposed an exact correspondence between such gauge theory operators
and string states on the Penrose limit of the $AdS_5 \times S^5$ geometry. Working solely within the deformed gauge theory
we use $\NN=1$ superspace methods, in a fashion first proposed in \italian,  to determine their exact
scaling dimensions for any value of the perturbing parameters and at strong t' Hooft coupling.
In general, these operators are not protected, since they do not fall
into short multiplets of the $SU(2,2|1)$ superconformal group and they obtain
anomalous dimensions as one moves away from the weakly coupled $\NN=4$ SYM point. Scaling dimensions
of such non-protected operators are expected (already at the $\NN=4$ point) to diverge at strong t' Hooft coupling
as $(g_{YM}^2 N)^{1/4}$, but as a special property of the large $R$-charge limit of \bmn\ they
approach a finite value at strong t' Hooft coupling.

With these operators at hand and following the spirit of the proposal in \bmn, we can further ask for a light-cone
worldsheet theory, whose spectrum reproduces the scaling dimensions we
found. Once the worldsheet theory has been determined,
we can further attempt to read off the dual string theory background. We find that such a
process does not result in
a unique background in the infinite R-charge limit. There is, however, a unique one which exhibits
supersymmetry
enhancement from sixteen to twenty-four
supersymmetries.

This reverse-engineering of a string theory
from data available in gauge theory would provide, in general, a very powerful
method for uncovering further examples of gauge-gravity duals
and one would like to have, if possible, a generic
prescription
to achieve it. In this paper we use the very special properties of the
correspondence proposed in \bmn;
in order to achieve a similar task in a more generic situation one would first have to understand better how
to extend this correspondence to finite R-charge and in cases without conformal invariance and/or no supersymmetry.

For the $\NN=4$ SYM theory at large R-charge, we should focus on the Penrose limit of $AdS_5 \times S^5$
\refs{\penrose,\onepapado,\twopapado,\threepapado}.
This limit leads to a maximally supersymmetric background with metric
\eqn\pp{ds^2=-4 dx^+ dx^- + \sum^8_{i=1} (dr^i dr^i -r^i r^i (dx^+)^2),}
and constant R-R 5-form flux
\eqn\flux{F_{+1234}=F_{+5678}=\rm{const}.}
One of the merits of this background is the exact solvability of the associated worldsheet theory
in the light-cone Green-Schwarz formalism, where it simply reduces to a sum of massive oscillators
\refs{\metsaev,\mettseyt}.
On the gauge theory side the Hilbert space of the $\NN=4$ SYM is
suitably truncated to states with large scaling dimension $\Delta \sim \sqrt{N}$ and large $U(1)_R$ $R$-charge
$J \sim \sqrt N$, while the difference $(\Delta - J)$
is kept fixed and small. A correspondence between such states and on-shell
states of string theory in the bulk pp-wave background was proposed by
Berenstein, Maldacena and Nastase (BMN)
in \bmn\ and as a check the scaling dimensions on both sides were computed and were found to agree.
Further checks of this correspondence (and beyond the planar limit)
were performed in
\refs{\italian,\staudacher\gross\motl\gursoy\staudachertwo\headrick-\beisert}.

We can obtain a whole moduli space of $\NN=1$ SYM theories by perturbing the $\NN=4$ Lagrangian
by a superpotential that breaks the $SU(4)_R$ $R$-symmetry group to a diagonal $U(1)_R$
under which all six of the Higgs fields are charged. This $U(1)_R$ is different from the one that was considered
in \bmn\ and for that reason it is useful to present a slight variant of that discussion for the $\NN=4$ theory.
We perform the Penrose limit of $AdS_5 \times S^5$ around the appropriate geodesic and repeat the BMN analysis to
rephrase the correspondence between string theory and gauge theory. We find that the resulting pp-wave limit
has a metric of the form
\eqn\ppmagnet{ds^2=-4 dx^+ dx^- + 4\mu y_1 dx_1dx^+ +4\mu y_2 dx_2 dx^+ - \mu^2 \vec{r}^2 (dx^+)^2 +d\vec{r}^2
+d\vec{y}^2 + d\vec{x}^2,}
and a 5-form field strength of the form
\eqn\ppmagform{F_5={\cal F}_5+*{\cal F}_5, \ \ {\cal F}_5 \sim \mu dx^+ \wedge dy^1 \wedge dx^1 \wedge dy^2 \wedge dx^2.}
$\mu$ is a mass parameter that can be scaled out through the rescaling $x^+ \rightarrow x^+/\mu$ and $x^- \rightarrow
\mu x^-$. In the rest of the paper it is set to one.
The Green-Schwarz light-cone worldsheet action includes four massive harmonic oscillators as in \bmn\ and
a Landau part that corresponds to the action of a charged particle moving in
the presence of a constant magnetic field. This action is again exactly solvable and the string spectrum is
known. In fact, after a suitable $x^+$-dependent change of coordinates the magnetic background of \ppmagnet\ transforms
into \pp\ \gomis. On the gauge theory side,
the Penrose limit restricts the $\NN =4$ SYM Hilbert space into
the same subsector as the one that appears in \bmn, but the R-charge assignments are now different. As a result, the
BMN correspondence involves at each level an infinite degeneracy.
On the string theory side this is the usual infinite degeneracy
of Landau levels.

The organization of this paper is the following.
In section 2, we discuss in detail the Penrose limit of interest and derive the resulting geometry at the $\NN=4$ point.
We consider string propagation on this geometry and review the associated string spectra. Then, we focus
on the gauge theory side and construct the string oscillators from the appropriate gauge invariant SYM
operators in the spirit of \bmn. This analysis is useful, because it clarifies some characteristics of the BMN correspondence
under a different R-charge assignment and it hints as to what may be
expected
to change or remain the same as we deform away from the $\NN=4$ point.
In section 3 we
briefly review the 2-complex parameter class of exactly marginal deformations of the $\NN=4$ SYM theory
that will be the main focus of our analysis.
This class of theories was introduced in \leigh\ and further studied in connection with
AdS/CFT in  \refs{\novel, \berestwo, \beresone, \fayya,\aharony} .
We proceed to determine the properties of
the BMN operators after the Leigh-Strassler deformations using $\NN=1$ superspace techniques.
We write down appropriate two-point functions of these operators and deduce their exact scaling dimensions
in a fashion similar to \italian. As a further check of this result, we perform a perturbative calculation to
verify in leading order that the scaling dimensions depend on the deforming parameters as expected.
In section 4 we use the available gauge theory data to reconstruct the
worldsheet action for string propagation in the Penrose limit
of the dual geometry and provide a detailed analysis of the supersymmetries
preserved by the associated pp-wave.
In section 5 we present our conclusions and suggest
directions for further research.

\newsec{A ``magnetic'' pp-wave limit of $AdS_5 \times S^5$ and its gauge theory dual}

\subsec{The Penrose limit}

Let us start with the $AdS_5 \times S^5$ metric
\eqn\adssmetric{ds^2=R^2(-dt^2 \cosh ^2 \rho + d\rho ^2+ \sinh ^2 \rho d\Omega_3 ^2 +
d\psi^2 \cos^2 \theta + d\theta^2 + \sin ^2 \theta d\Omega_3'^2 )}
and write explicitly the solid angle $d\Omega_3'^2$ in $S^5$ as
\eqn\solids{d\Omega_3'^2=\cos^2 \phi_1 d\phi_2^2 + d\phi_1^2 +\sin^2 \phi_1 d\phi_3^2.}
In this parametrization, $S^5$ is given in terms of the five coordinates $(\psi,\theta,\phi_1,
\phi_2,\phi_3)$. There are three obvious $U(1)$ isometries and they have to do with translations of the
coordinates $\psi,\phi_2$ and $\phi_3$.
On the gauge theory side
each of them is in one-to-one correspondence
with a $U(1)_R$ that rotates one of the three complex Higgs fields of the $\NN =4$ theory.
We denote them as $\Phi^1,\Phi^2$ and $\Phi^3$.
We make the correspondence
\eqn\jx{\Phi^1 \leftrightarrow J_{\Phi^1} = -i \d _{\psi},}
\eqn\jy{\Phi^2 \leftrightarrow J_{\Phi^2} = -i \d _{\phi_2},}
\eqn\jz{\Phi^3 \leftrightarrow J_{\Phi^3} = -i \d _{\phi_3}.}

In general, we would like to consider an arbitrary linear superposition of
the three $U(1)$ isometries  under which
the complex fields $\Phi^1,\Phi^2$ and $\Phi^3$ have charges $Q_1,Q_2$ and $Q_3$ respectively.
The Penrose limit will be taken along a null geodesic associated to this isometry. For that purpose
we introduce an angular coordinate $\omega'$ given by
\eqn\omegadef{-i \d _{\om '} \equiv -i (Q_1 \d _{\psi}+Q_2 \d_{\phi_2}+Q_3 \d_{\phi_3})}
and we suitably rescale it to get a new coordinate $\omega$ with periodicity $2\pi$. Independently of the
charges $Q_1, Q_2$ and $Q_3$, we can
always write $\omega =\frac{\psi+\phi_2+\phi_3}{3}$
and the charge of every complex Higgs field, as
measured by the current $-i\partial_{\omega}$, is one.

The geodesic of interest is given by
\eqn\nullgeo{t=\omega, \ \rho=0, \ \theta=\theta_0, \ \phi_1=\frac{\pi}{4}, \ \psi=
\phi_2=\phi_3=\omega,}
with $\theta_0=\arccos (1/\sqrt{3} )$. Indeed, a simple substitution of these values in
\adssmetric\ gives the null geodesic condition
\eqn\geozero{ds^2=R^2\bigg(-dt^2+d\omega^2\bigg)=0.}
In order to focus on the geometry of the neighborhood of this geodesic we introduce new coordinates
\eqn\xplus{x^+=\frac{1}{2}(t+\omega),}
\eqn\xminus{x^-=\frac{R^2}{2}(t-\omega)}
and perform the rescaling
\eqn\rescalingone{\rho=\frac{r}{R}, \ \theta=\theta_0+\frac{y_1}{R}, \ \phi_1=\frac{\pi}{4}+
\sqrt{\frac{3}{2}} \frac{y_2}{R}, \ \psi=\omega-\sqrt{2} \frac{x_1}{R},}
\eqn\rescalingtwo{\phi_2=\omega+\frac{1}{\sqrt{2}} \frac{x_1-\sqrt{3}x_2}{R}, \
\phi_3=\omega+\frac{1}{\sqrt{2}} \frac{x_1+\sqrt{3} x_2}{R},
}
taking the $ R \rightarrow \infty$ limit.
The numerical factors have been inserted for later convenience.

Expanding each expression in \adssmetric\ up to second order in $1/R^2$ gives the pp-wave metric
\eqn\ppwavemetric{ds^2=-4 dx^+ dx^- - r^2 (dx^+)^2 + \sum_{i=1}^4 dr^i dr^i +
\sum _{a=1,2}(dy_a^2+ dx_a^2 + 4 y_a dx_a dx^+).}
The full solution is also supported by the constant 5-form flux of eq.\ppmagform. Following \zayas\
we will hereafter refer to this background as the magnetic pp-wave limit of $AdS_5 \times S^5$. It is
a maximally supersymmetric background with 32 supersymmetries and its gauge theory dual is
a suitable truncation of the $\NN =4$ SYM.
This truncation is independent of the choice of the $U(1)_R$ and
therefore it is not different from the one that appears in \bmn.
It is worth noticing that the same pp-wave background also appears in \refs{\gomis,\zayas,\mukhi}, where
the Penrose limit was taken on $AdS^5 \times T^{1,1}$. The gauge theory dual in that case is an
$\NN =1$ $SU(N) \times SU(N)$ SYM with a pair of bifundamental chiral multiplets $A_i$ and $B_i$
transforming in the $(N, \bar{N})$
and $(\bar{N}, N)$ representation of the gauge group. The fact that it can also be obtained from
$AdS_5 \times S^5$ in the fashion that we discuss here was also mentioned in \zayas.

The correspondence between the light-cone momenta $p^-$ and
$p^+$ on the string theory side and the scaling dimensions and $R$-charges on the gauge theory side
works in the following way
\eqn\pplus{2p^-=-p_+=i\d_{x^+}=i(\d_t+\d_{\omega})=\Delta - J}
and
\eqn\pminus{2p^+=-p_-=i\d_{x^-}=\frac{1}{R^2}i(\d_t-\d_{\omega})=\frac{1}{R^2}(\Delta+J).}
$R$ is the radius of $AdS_5$ and we have set
\eqn\jcharge{
J=-i\d_{\omega}=\frac{1}{Q_1}R_1+\frac{1}{Q_2}R_2+\frac{1}{Q_3}R_3.}
For each $i=1,2,3$, $R_i$ is a $U(1)$ generator under which only $\Phi^i$ is charged and the charge is $Q_i$.

In the limit under consideration $R \rightarrow \infty$. Since we only keep the states with finite $p^+$ it is necessary
to take the familiar scaling $\Delta, J \sim R^2 \sim \sqrt{N}$. As a result, on the
gauge theory side we must take the $N \rightarrow \infty$ limit
keeping the Yang-Mills coupling fixed and small and focus on operators
with large $R$-charge $J \sim \sqrt{N}$ and small and fixed
$\Delta-J$. Such operators were introduced in \bmn\ and we re-discuss them in the magnetic pp-wave context
in section 2.3.

\subsec{String propagation on magnetic pp-waves}

The gauge-fixed light-cone bosonic string action for the background \ppwavemetric\ is \zayas
\eqn\bosonaction{S=\frac{1}{2\pi \alpha'} \int d\tau \int_0^{2\pi \alpha' p^+} d\sigma \bigg (
\frac{1}{2} \d_a \vec{r} \d^a \vec{r} - \frac{1}{2} r^2
+\frac{1}{2} \d_a \vec{x} \d^a \vec{x}
+\frac{1}{2}\d_a \vec{y} \d^a \vec{y} -2 \vec{x} \d_{\tau} \vec{y} \bigg).}
There are several terms contributing to this action. There are four massive oscillators labeled
by the 4-dimensional vector $\vec{r}$ and two identical decoupled Landau actions involving the coordinates
$(x_1,y_1)$ and $(x_2,y_2)$. Each of them is precisely the action of a 2-dimensional charged particle
moving in a constant magnetic field. It is convenient to rewrite the $x-y$ part of the action by performing
the rotation
\eqn\rotation{x_a=-\frac{1}{\sqrt{2}}(\hat{x}_a+\hat{y}_a), \ \ y_a=\frac{1}{\sqrt{2}}(\hat{x}_a-\hat{y}_a).}
Up to a total derivative term that can be dropped the action takes the form
\eqn\xyrotated{S_{xy}=\frac{1}{2\pi \alpha'} \int d\tau \int_0^{2\pi \alpha' p^+}
d\sigma \bigg( \frac{1}{2} \d_a \vec{\hat x} \cdot \d^a \vec{\hat x} +\frac{1}{2} \d_a \vec{\hat y} \cdot
\d^a \vec{\hat y}-
\vec{\hat x} \cdot \d_{\tau} \vec{\hat y} + \vec{\hat y} \cdot \d_{\tau} \vec{\hat x} \bigg)}
and from now on we drop the $\hat{}$ notation. This action and the associated spectrum have also appeared in
the context of the Penrose limit of $AdS_5 \times T^{1,1}$ in \gomis. For completeness, in the rest of this subsection
we review the spectra that were obtained there.

The spectrum of the $r^i$ part of the light-cone Hamiltonian reads
\eqn\rhamiltonian{\HH_r=\sum_{n=-\infty}^{\infty} N_n^{(r)} \sqrt{1+\bigg( \frac{n}{\alpha' p^+} \bigg)^2}.}
There are four kinds of oscillators contributing to the level $N_n^{(r)}$ and we denote them
as $a_n^i$, for $i=1,2,3,4$. We use the notation of \bmn, so $n>0$ label the left movers and $n<0$ label the
right movers.

For the $x-y$ part of the action the light-cone Hamiltonian breaks up into four parts
\eqn\xyhamiltonian{\HH_{xy}=\sum_{n=-\infty}^{\infty} \sum_{a=1,2} \bigg[
N_n^{(b^a)} \bigg( \sqrt{1+\bigg( \frac{n}{\alpha' p^+} \bigg)^2}+1 \bigg) +
N_n^{(\bar{b}^a)} \bigg( \sqrt{1+\bigg( \frac{n}{\alpha' p^+} \bigg)^2}-1 \bigg) \bigg].}
Four types of oscillators contribute to each of the above terms.
The oscillators $(b_n^1,\bar{b}_n^1)$ originate from the $(x_1,y_1)$ part of the Lagrangian and
contribute to the levels $N_n^{b^1}$ and $N_n^{\bar{b}^1}$ respectively and the
oscillators $(b_n^2,\bar{b}_n^2)$ contribute to the levels $N_n^{b^2}$ and $N_n^{\bar{b}^2}$.

These spectra can be derived by straightforward calculation, or they can be deduced from
the following slightly different point of view \gomis. After the change of variables \rotation,
we introduce the complex coordinates $z_a=x_a+iy_a$ and we bring
the metric \ppwavemetric\ into the form
\eqn\ppzmetric{ds^2=-4dx^+ dx^- -r^2 (dx^+)^2+ \sum_{i=1}^4 dr^i dr^i + \sum_{a=1,2} (dz_a d\bar{z_a} +
i(\bar{z}_a dz_a-z_ad\bar{z}_a)dx^+).}
This background can be transformed into the maximally supersymmetric
pp-wave solution of \bmn\ if we perform the $x^+$-coordinate dependent $U(1) \times U(1)$ rotation
\eqn\xplusrotation{z_a=e^{ix^+} w_a, \ \ \bar{z}_a=e^{-ix^+}\bar{w}_a.}
In view of \pplus\ this translates to
\eqn\newdj{\eqalign{\Delta-J&=i\d_{x^+}|_{z_a}
\cr
=i\d_{x^+}|_{w_a}+\sum_a (w_a{\d_{w_a}}-\bar{w}_a & \d_{\bar{w}_a} )={(\Delta-J)}_{S^5}+J_1+J_2,}}
where $J_1$ and $J_2$ are $U(1)$ rotation charges in the $(w_1,\bar{w}_1)$ and $(w_2,\bar{w}_2)$ transverse planes
respectively.

The spectra of eqs.\rhamiltonian,\xyhamiltonian\ can be reproduced from \newdj\ by noticing that the bosonic
oscillators have the following $J_1,J_2$ charges
\eqn\bosjcharges{\eqalign{a^i_n& \ \ J_1=J_2=0, \ \ \ \ i=1,2,3,4
\cr
b^1_n& \ \ J_1=1, J_2=0,
\cr
\bar{b}^1_n& \ \ J_1=-1, J_2=0,
\cr
b^2_n& \ \ J_1=0, J_2=1,
\cr
\bar{b}^2_n& \ \ J_1=0, J_2=-1.}}

The fermionic oscillator contributions to the light-cone Hamiltonian $p^-$ can be
similarly deduced from \newdj\ by looking at the
$U(1) \times U(1)$ charges carried by the $SO(8)$ spinor ${\bf 8}_s$ under the $SU(2) \times SU(2)
\times U(1) \times U(1)$ into which $SO(8)$ has been broken \gomis
\eqn\eights{{\bf 8}_s \rightarrow ({\bf 2}, {\bf 1})_{(1/2,1/2)} \oplus ({\bf 2}, {\bf 1})_{(-1/2,-1/2)}
\oplus ({\bf 1}, {\bf 2})_{(1/2,-1/2)} \oplus ({\bf 1}, {\bf 2})_{(-1/2,1/2)}.}
We get the spectra
\eqn\spinors{\eqalign{
S^{\alpha ++}_n& \ \ 2p^-=\sqrt{1+\bigg( \frac{n}{\alpha' p^+} \bigg)^2} +1,
\cr
S^{\alpha --}_n& \ \ 2p^-=\sqrt{1+\bigg( \frac{n}{\alpha' p^+} \bigg)^2} -1,
\cr
S^{\dot{\alpha} +-}_n& \ \ 2p^-=\sqrt{1+\bigg( \frac{n}{\alpha' p^+} \bigg)^2},
\cr
S^{\dot{\alpha} -+}_n& \ \ 2p^-=\sqrt{1+\bigg( \frac{n}{\alpha' p^+} \bigg)^2},}}
which, as expected, turn out to be identical to the bosonic ones.

Notice that the action of the bosonic zero mode
oscillators $\bar{b}^{1}_0$ and
$\bar{b}^{2}_0$, as well as the
action of their fermionic superpartners $S^{\alpha --}_0$ has no effect on the light-cone energy.
As a result, the spectrum exhibits an infinite degeneracy. The degenerate states are obtained
by the action of an arbitrary number of the above zero mode oscillators on the vacuum.
This degeneracy is familiar, since the worldsheet action contains two decoupled Landau parts,
which describe a charged particle moving in the presence of a constant magnetic field in ${\bf R}^2
\times {\bf R}^2$. This system is known to have an infinite degeneracy of states labeled by the
angular momentum of the charged particle.

In the next section we discuss how these bulk characteristics manifest themselves on the dual
gauge theory.

\subsec{The gauge/string correspondence}

Now we would like
to discuss the correspondence between the string oscillator states of the previous section and appropriate
operators in the dual $\NN =4$ SYM theory. Following \bmn\ we are interested in
the large $N$ limit with $g_{YM}^2$ kept fixed and small. We work in the planar limit and
examine single trace operators, which we categorize by their $\Delta -J$ value. As in the usual BMN
limit there exists a very interesting finite $J$ version of these operators
 \refs{\andryzhov, \beisert},
which we do not discuss in this paper.

We begin with single trace operators of $\Delta -J=0$. There is an infinite
number. Any traceless operator of the form $\tr [\Phi^1...\Phi^2...\Phi^3...]$ containing $J$ symmetrized insertions
of the $\Phi^1,\Phi^2$ or $\Phi^3$ fields has $\Delta -J=0$.
Each of them is an $\NN=4$ chiral primary and its scaling dimension is protected by supersymmetry.

In order to construct the correspondence of SYM operators with string oscillator states, it is perhaps natural to
single out a specific linear superposition of the Higgs fields associated to the $U(1)_R$ generator
$J$ that appears in \jcharge. We choose the diagonal superposition
\eqn\defomega{\Omega = \frac{1}{\sqrt{3}} (\Phi^1+\Phi^2+\Phi^3).}
In the language of \bmn\ we propose the correspondence
\eqn\omegatrace{\frac{1}{\sqrt{J} N^{J/2}}\tr [\Omega^J] \leftrightarrow |0,p^+;\sigma_{\Omega} \rangle_{l.c.},}
where $\sigma_{\Omega}$ is a formal parameter that denotes
a particular state of the infinitely degenerate
light-cone vacuum space. The factor of the l.h.s. is such that the normalization of the two point function
is one.

To obtain the rest of the $\Delta -J=0$ operators we act on the above vacuum with an
arbitrary number of the zero mode oscillators $\bar{b}_0^{1},\bar{b}_0^{2}$. Since they have no effect on the
light-cone energy, these oscillators should be associated again to linear combinations of
the Higgs fields $\Phi^1,\Phi^2$ and $\Phi^3$. We choose the two linear combinations that are
orthogonal to $\Omega$ and propose the correspondence
\eqn\czeroone{\bar{b}_0^1 \leftrightarrow \Psi^1 = \frac{1}{\sqrt{2}}( \Phi^2+\Phi^3-2\Phi^1)}
and
\eqn\czerotwo{\bar{b}_0^2 \leftrightarrow \Psi^2 = \frac{1}{\sqrt{6}} (\Phi^3-\Phi^2).}

It is clear that the above correspondence between operator insertions and string oscillators is by no means unique.
Any $SU(3)$ rotated basis of Higgs fields could equally well be assigned to the same string oscillators.
This lack of uniqueness is also manifest on the arbitrary choice of the state
$|0,p^+;\sigma_{\Omega} \rangle_{l.c.}$ on the r.h.s. of \omegatrace.

With the above correspondence
the action of the zero mode oscillators $\bar{b}_0^{a \dagger}$ ($a=1,2$) on the light-cone vacuum
\omegatrace\ can be translated in the SYM language as follows. For each $\bar{b}_0^{a \dagger}$ we are instructed to
make an insertion of $\Psi^a$ and then sum over all possible orderings.
This is the same as acting on $\tr [\Omega^J]$ with the operator $\sum_{l=1}^J(\Omega \Psi^a
\frac{\d}{\d \Omega})_l$, where we use the notation $(...)_l$ to denote that the operator in parenthesis acts on
the $l$th insertion of the trace.
For example,
\eqn\exampleczero{\frac{1}{\sqrt{J}}\sum_l \frac{1}{\sqrt{J} N^{J/2+1/2}}{\rm Tr}[\Omega^l \Psi^a \Omega^{J-l-1}]
\leftrightarrow \bar{b}_0^{a\dagger} |0,p^+;\sigma_{\Omega}
\rangle_{l.c.}.}
Repeated action of these zero modes creates the anticipated Landau degeneracy of the vacuum, which becomes
infinite in the $J \rightarrow \infty$ limit.

For the operators with $\Delta -J=1$ we can say the following. There are {\it twelve} bosonic operators of this type,
$D_i \Omega$, $D_i \Psi^1$ and $D_i \Psi^2$ and they are expected
to match the four zero mode oscillators $a_0^i$ for $i=1,2,3,4$.
This correspondence works by associating
\eqn\ai{{a^i}^{\dagger} \leftrightarrow \sum_{l} (\Omega D_i)_l \ \ \ i=1,2,3,4.}
$D_i$ denotes the gauge covariant derivative with respect to the spacetime coordinates of ${\bf R}^4$ where
the dual $\NN =4$ gauge theory lives.
More precisely, whenever we act on the vacuum $|0,p^+;\sigma_{\Omega} \rangle
_{l.c.}$ of eq.\omegatrace\ with the oscillator ${a^i}^{\dagger}$, we are instructed to add an insertion of
$D_i \Omega$ on the gauge theory operator $\tr [\Omega^J]$ and then sum over all possible orderings, e.g.
\eqn\aivac{\frac{1}{\sqrt{J}}\sum_{l=0}^J \frac{1}{\sqrt{J} N^{J/2+1/2}}{\rm Tr}[\Omega^l D_i\Omega \Omega^{J-l}]
\leftrightarrow {a_0^i}^{\dagger} |0,p^+;\sigma_{\Omega} \rangle_{l.c.}.}
Acting on a different vacuum state of the same light-cone energy, e.g.\ acting on
${\bar b}_0^{a\dagger}|0,p^+;\sigma_{\Omega} \rangle_{l.c.}$, also amounts to a similar insertion of
$D_i \Omega$ or $D_i \Psi^a$.
We insert $D_i \Omega$ if a position is initially occupied by $\Omega$ and $D_i \Psi^a$
if the position is initially occupied by $\Psi^a$.
This rule is a consequence of the fact that the
state ${a^i_0}^{\dagger}{\bar b}_0^{a\dagger}|0,p^+;\sigma_{\Omega} \rangle_{l.c.}$
can also be written as ${\bar b}_0^{a\dagger}{a^i_0}^{\dagger}|0,p^+;\sigma_{\Omega}
\rangle_{l.c.}$.

Finally, we have to consider insertions of the $\Delta-J=2$
operator $\bar{\Psi}^a$. From the string spectrum \xyhamiltonian\
it is apparent that such insertions correspond to the action of the
zero mode oscillators ${b_0^a}^{\dagger}$, which
increase the light-cone Hamiltonian by 2.
It is therefore natural to make the identification
\eqn\bzerovac{\frac{1}{\sqrt{J}}\sum_l \frac{1}{\sqrt{J} N^{J/2+1/2}}{\rm Tr}[\Omega^l  (\bar{\Psi}^a) \Omega^{J-l}]
\leftrightarrow {b_0^a}^{\dagger} |0,p^+;\sigma_{\Omega} \rangle_{l.c.}.}

The above correspondence also extends nicely to the fermionic zero mode oscillators \spinors.
The relevant SYM operators follow easily from the bosonic ones
by supersymmetry. We have
\eqn\fermioniccor{\eqalign{
{\rm \bf gauge \ theory\ ferm}&{\rm \bf ionic \ operators} \ \ \leftrightarrow \ {\rm \bf fermionic \ string
\
\bf oscillators}
\cr
&\bar{\lambda}^{\dot{\alpha}+-} \ \ \ \ \ \ \ \ \ \ \ \ \ \ \ \ \ \ \ \ \ \ \ \ \ \ \ \ \ \ \ \ \
\ \ \ S^{\dot{\alpha}+-},
\cr
&\bar{\lambda}^{\dot{\alpha}-+} \ \ \ \ \ \ \ \ \ \ \ \ \ \ \ \ \ \ \ \ \ \ \ \ \ \ \ \ \ \ \ \ \
\ \ \ S^{\dot{\alpha}-+},
\cr
&\bar{\psi}^1  \ \ \ \ \ \ \ \ \ \ \ \ \ \ \ \ \ \ \ \ \ \ \ \ \ \ \ \ \ \ \ \ \ \ \ \ \
\ \ \  S^{1++},
\cr
&\bar{\psi}^2   \ \ \ \ \ \ \ \ \ \ \ \ \ \ \ \ \ \ \ \ \ \ \ \ \ \ \ \ \ \ \ \ \ \ \ \ \
\ \ \  S^{2++},
\cr
&\psi^1           \ \ \ \ \ \ \ \ \ \ \ \ \ \ \ \ \ \ \ \ \ \ \ \ \ \ \ \ \ \ \ \ \ \ \ \ \
\ \ \ S^{1--},
\cr
&\psi^2 \ \ \ \ \ \ \ \ \ \ \ \ \ \ \ \ \ \ \ \ \ \ \ \ \ \ \ \ \ \ \ \ \ \ \ \ \
\ \ \ S^{2--}.
}}
$\bar{\lambda}$ denotes the right-handed gauginos. There are 8 such components. Each of them has a
definite charge ($\pm 1/2$) under the two ``Landau'' $U(1)$'s into which $SO(4) \subset SO(6)_R$ has been broken.
The $+/ -$ superscripts denote the components of the gauginos with charges $\pm 1/2$ respectively.
$\psi^a$ for $a=1,2$ are the fermionic superpartners of the bosons $\Psi^a$.

For the higher excited modes of the string the correspondence works exactly as in \bmn.
The action of any excited oscillator is expressed in the SYM language by the insertion of the corresponding
field multiplied by a position dependent phase, e.g.
\eqn\excitedstate{\frac{1}{\sqrt{J}}\sum_l \frac{1}{ N^{J/2+1}}{\rm Tr}[\Psi^a \Omega^l  \Psi^b \Omega^{J-l}]
e^{\frac{2\pi i nl}{J}}
\leftrightarrow {\bar{b}}_n^{b\dagger}\bar{b}_{-n}^{a\dagger} |0,p^+;\sigma_{\Omega} \rangle_{l.c.}.}
The details of this construction are precisely the same as in \bmn\ and we will not discuss them further.

In conclusion, we rephrased the BMN correspondence at the $\NN=4$ SYM fixed line for a diagonal $U(1)_R$
choice. We did not go into much detail, because the essence of the correspondence is expected to be
independent of this choice and in particular, it should be
easy to translate all the checks and extensions of the correspondence at finite $J$ in the language of this section.
Furthermore, it is natural to expect that this same BMN correspondence also persists when we deform away from
the $\NN=4$ fixed line. The goal of the next section is to determine the effect of the deformation on the
BMN operators.

\newsec{$\NN=1$ superconformal theories and BMN operators}

\subsec{Exactly marginal deformations of the $\NN=4$ SYM theory}

After this long parenthesis on magnetic pp-waves, we are now ready to proceed with the analysis of
the Leigh-Strassler deformations of the $\NN=4$ SYM theory.
The four-dimensional $\NN=4$ $SU(N)$ SYM theory can be expressed in the language of $\NN=1$ supersymmetry
in terms of a vector multiplet $V$ and three chiral multiplets\foot{In this section  $\Phi^i$, $\Omega$ and $\Psi^a$
denote full $\NN=1$ superfields and they should not be confused
with the bosonic bottom components of the previous section.}
$\Phi^i$, $i=1,2,3$. In addition to the usual kinetic terms
of the $\NN=1$ theory one is also instructed to add a superpotential of the form
\eqn\superpotential{W=g' {\rm Tr}([\Phi^1,\Phi^2]\Phi^3).}
In this $\NN=1$ language only an $SU(3) \times U(1)$ subgroup of the full $SU(4)_R$ $R$-symmetry group is
manifest. $SU(3)$ is the group that rotates the chiral superfields $\Phi^i$.
At the $\NN=4$ point the superpotential coupling $g'$ is directly related to the Yang-Mills coupling and in our conventions
$g'=\sqrt{2} g_{YM}$. To set our notation straight we write
the full $\NN=4$ action as
\eqn\fullfour{\eqalign{
\SS = \tr & \bigg ( \int d^4 \theta e^{-gV} \bar{\Phi}_i e^{gV} \Phi^i +
\frac{1}{2g^2} \bigg [ \int d^4 x d^2 \theta W^{\alpha} W_{\alpha} +
\int d^4 x d^2 \bar{\theta} \bar{W}^{\dot{\alpha}}\bar{W}_{\dot{\alpha}} \bigg ] +
\cr
&+ \frac{g'}{3!} \int d^4 x d^2 \theta \epsilon_{ijk} \Phi^i [\Phi^j,\Phi^k] -
\frac{g'}{3!} \int d^4 x d^2 \bar{\theta} \epsilon^{ijk} \bar{\Phi}_i
[\bar{\Phi}_j, \bar{\Phi}_k ] \bigg)
}}
and by definition we always set $g=\sqrt{2}g_{YM}$.
Notice the explicit distinction between the superpotential coupling $g'$ and the vector superfield coupling $g$. At
the $\NN=4$ fixed line we have $g=g'$ but
this relation is modified as we deform away and in general
we need to differentiate between the two couplings.

Since the $\NN=4$ theory is conformal for any value of the complex
coupling $\tau=\frac{\theta}{2\pi}+\frac{4\pi i}{g_{YM}^2}$, the deformation that changes this value is obviously
exactly marginal. It is also known, however, that for $N \geq 3$ the $\NN=4$ theory has additional exactly
marginal perturbations \leigh. Classically, one possibility is given by the superpotential
\eqn\sympotential{W=h_{ijk} {\rm Tr} (\Phi^i \Phi^j \Phi^k),}
with ten symmetric coefficients $h_{ijk}$.
Another one is the superpotential \superpotential\ with any (complex)
coefficient $g'$. For the first class,
it is known \refs{\aharony,\bigads,\leigh} that only a two-complex parameter
subset of them is exactly marginal on the quantum level. The resulting superpotential can be written as
\eqn\twopotential{W_{def}=h_1 {\rm Tr}(\Phi^1 \Phi^2 \Phi^3 + \Phi^1 \Phi^ 3 \Phi^2) + h_2 {\rm Tr}({(\Phi^1)}^3+{(\Phi^2)}^3+
{(\Phi^3)}^3),}
in terms of two complex coefficients $h_1,h_2$. These particular deformations preserve a $Z_3 \times Z_3$
symmetry given by the transformations $\Phi^1 \rightarrow \Phi^2$, $\Phi^2 \rightarrow \Phi^3$, $\Phi^3 \rightarrow \Phi^1$
and $\Phi^1 \rightarrow \Phi^1$, $\Phi^2 \rightarrow \omega \Phi^2$, $\Phi^3 \rightarrow \omega^2 \Phi^3$.
$\omega$ is a cubic root of unity. The second $Z_3$ prevents any mixing between the chiral operators $\Phi^i$ and
the first can be used to show that they all have the same anomalous dimension $\gamma(\tau,g',h_1,h_2)$.
The beta functions are restricted by non-renormalization theorems to be
proportional to this anomalous dimension and the constraint
\eqn\constraint{\gamma(g,g',h_1,h_2)=0}
gives a 3-complex dimensional surface of fixed points. For simplicity, we
set the theta angle to zero.

The analytic form of this surface is only known up to first
order in perturbation theory \refs{\aharony, \razamattwo, \razamat}.
Notice that for generic points in this moduli space
the coefficient $g'$ is not necessarily
equal to the $\NN = 4$ value $g=\sqrt{2} g_{YM}$. It turns out that
the large $R$-charge limit, on which we base our analysis, probes
a neighborhood of this moduli space around the strong 't-Hooft coupling point. Thus, for
later considerations it is convenient to write $g'$ as $g'=g+h_0$, with $h_0$ complex.
At the end of the day, our results on the anomalous dimensions
of the BMN operators will be expressed in terms
of the three independent couplings $g, h_1$ and $h_2$.

The conclusion of this short introduction is that for fixed $g$
there are basically two exactly marginal deformations away from the $\NN=4$ fixed line and they correspond to
the superpotential \twopotential. On the supergravity side this deformation can be
identified at first order with part of the KK scalar mode in the
$\bf 45$ of $SO(6)$ \refs{\novel,\romans}. This scalar corresponds to the second two-form harmonic
$Y^I_{[\alpha,\beta]}$ in
the expansion of the complex antisymmetric two-form
$A_{\alpha,\beta}$
with components along the five-sphere.
The effect of the deformation in supergravity has been analyzed perturbatively in the deformation parameters
in \refs{\novel,\aharony} and is expected to be
a warped fibration of AdS$_5$ over a deformed $\tilde{S}^5$ in the presence
of 3-form and 5-form fluxes.
An interesting class of supergravity solutions
of this type was also obtained in \fayya. These solutions, however, appear to be singular and their exact relation to
the deformation superpotential \twopotential\ is not clear.

\subsec{BPS and near-BPS operators}

In section 2 and in the context of a ``magnetic'' Penrose limit of $AdS_5 \times S^5$ we considered a class of
large R-charge operators of the $\NN=4$ SYM theory, which were obtained from the operator
\eqn\groundoperator{{\Pi}_J \equiv \frac{1}{\sqrt{J} N^{J/2}}\tr [\Omega^J]}
by insertions of the fields
$D_i \Omega$, $\Psi^a$ and $\bar{\Psi}^a$ ($i=1,2,3,4$ and $a=1,2$) with or without
position dependent phases.
Without such phases the resulting symmetrized operators are 1/2-BPS.
They are protected operators of the $\NN=4$ theory because they belong to short multiplets of
the $SU(2,2|4)$ superconformal group\foot{More specifically,
they are protected because they belong to short multiplets that cannot
combine to form long multiplets after the $\NN=4$ interaction is turned on.
See e.g.\ \osborntwo\ for a recent discussion on this point.} .

Alternatively, we can ask in what sense they are protected
from an $\NN=1$ point of view. Generically an $\NN=4$ short
multiplet can break into $\NN=1$ short and long multiplets and
it is not immediately obvious how the $\NN=4$ protection manifests itself
in the $\NN=1$ formalism. This question is even more important and instructive
in anticipation of the Leigh-Strassler deformation that
breaks the $\NN=4$ supersymmetry down to $\NN=1$. We need to know what remains protected even after the
deformation.
The $\NN=1$ of interest is
the one that is preserved by the Leigh-Strassler deformations, i.e. one under which all three Higgs fields
have equal R-charge 2/3.

Let us first see what happens along the $\NN=4$ fixed line from an $\NN=1$ point of view.
${\Pi}_J$ is protected, because it
is an $\NN=1$ chiral primary
operator and obeys the BPS condition $\Delta = J$. The same is also
true for the operators that arise when we include symmetrized insertions of the fields $\Psi^a$.
Insertions of the fields
$D_i \Omega$ lead to descendants of ${\Pi}_J$ and they are also protected. The
remaining operators are those
with $\bar{\Psi}^a$ insertions. Every such insertion has $\Delta - J=2$ at weak coupling and clearly
does not produce an $\NN=1$ chiral field. Nevertheless, the resulting operator is still $\NN=1$ protected,
because it belongs to another type of short multiplet of $SU(2,2|1)$ and in $\NN=1$ notation it
is known as a semi-conserved superfield (see, for example, \dalaras).
Semi-conserved superfields $L$ obey the condition\foot{$D_{\alpha}$
and $\bar{D}_{\dot{\alpha}}$ are the usual superspace
covariant derivatives. In what follows, we work in $\NN=1$ superspace and
adopt the notations of \wb.}
\eqn\semiconserved{
\bar{D}^2 L = 0.}
Using the $\NN=4$ SYM equations of motion one can easily verify that the corresponding superfields
with $\bar{\Psi}^a$ insertions indeed satisfy this condition.

On the other hand, operators
with the above insertions and position-dependent phases are not protected, because
the insertions are not symmetrized. For example, operators of the type
\eqn\exampleop{
\sum_l e^{\frac{2\pi i n}{J}} \tr [\Psi^a \Omega^l \Psi^b \Omega^{J-l} ]
}
have $\Delta-J=0$ at weak coupling and they may seem to be chiral and hence protected. This, however, is not correct,
because one can use the $\NN=4$ SYM equations of motion to symmetrize this operator. In the process extra
terms appear and they turn out to be descendants of non-protected operators.
A similar reasoning can also be applied to other non-symmetrized operators.

Once we deform the $\NN=4$ SYM action by the superpotential \twopotential\ at a generic point of the moduli
manifold \constraint\ many of
the above statements about the 0-level
BPS operators change. As we verify explicitly in the next section,
the deformation modifies the
$\NN=4$ equations of motion and the previously protected operators acquire nonzero anomalous
dimensions. For example,
it is easy to check that \semiconserved\ breaks down away from the $\NN=4$ point and
operators with $\bar{\Psi}^a$ insertions no longer remain semi-conserved in the deformed theories.
Similarly, the previously symmetrized chiral operators with $\Psi^a$ insertions
acquire anomalous dimensions and they are not protected against
the $\NN=4$-breaking deformations. These anomalous dimensions are computed in the next section
using the technology of \italian\ and they are verified independently to leading order in perturbation
theory in section 3.3.

Only one operator remains protected and continues to
have $\Delta-J=0$. This is $\Pi_J$.
The vanishing of its anomalous dimension is synonymous to the condition \constraint\ that guarantees the
presence of superconformal invariance in the deformed theory.
As a result, we see that the effect of the deformation is to lift the infinite Landau degeneracy of the $\NN=4$ point
and retain a single vacuum state represented on the gauge theory side by the operator $\Pi_J$. Such a vacuum state
with vanishing light-cone energy
should also be expected from the supersymmetry of the dual
background.

Another aspect of this picture is the following.
We have concentrated our attention on the BMN operators that can be obtained from $\Pi_J$ by appropriate insertions of
other fields and worked mainly in a ``dilute gas'' approximation. In doing so, we break the
$Z_3$ symmetry that permutes
the three adjoint chiral superfields and the ``vacuum'' operators
$\tr [{(\Psi^1)}^J]$ and $\tr [{(\Psi^2)}^J]$ remain at ``infinite distance'' from the operator $\Pi_J$,
i.e.\ they result from infinite insertions. This seems to be inconsequential for the BMN correspondence at the
$\NN=4$ point, because of the infinite Landau degeneracy,
but it is perhaps a little puzzling for the BMN correspondence after the deformation. These operators have similar
properties as $\Pi_J$ and they continue to
have $\Delta-J=0$ throughout the moduli space. In order to obtain them from
$\Pi_J$ we have to start adding insertions that
increase the total $\Delta-J$
\foot{For the type of $\Delta-J$ values that we find after the deformation, see for example Table 1.}
and it is not completely obvious how we can recover an operator
with $\Delta-J=0$.
The key point has to be that after several insertions the ``dilute gas'' approximation starts breaking down and
one has to be more careful on the derivation of the scaling dimensions. This process is also obscured by the
fact that we have to add an infinite number of insertions and this is not something completely well-defined.

\subsec{Exact scaling dimensions in superspace formalism}

In order to calculate the anomalous dimensions of the above operators, we would like to determine the
appropriate two-point functions. The authors of \italian\ performed a similar calculation at the $\NN=4$ point by
working in superspace formalism and using the constraint imposed by the equations of motion of the theory.\foot{A
similar calculation for $\NN=2$ superconformal gauge theories based on ADE quiver diagrams was performed
in \oz.}
Following their example, we consider the operators\foot{Gauge invariance
demands that the operator $\UU_J^a$ should be written as
$\sum_l e^{il \varphi} \Omega^l e^{-gV} \bar{\Psi}^a e^{gV} \Omega^{J-l}$. For our purposes, however, it is
enough to work with the assumption that $V=0$.}
\eqn\opu{\UU_J^a= \sum_l e^{il \varphi} \Omega^l \bar{\Psi}^a \Omega^{J-l}
}
and
\eqn\opo{\OO_J^a=\sum_l e^{il \varphi} \Omega^l \Psi^a \Omega^{J-l},
}
for $a=1,2$ and $\varphi=\frac{2\pi n}{J}$. The actual operators that appear
in the BMN construction are traced gauge invariant operators of the type
\eqn\opexample{
\sum_l e^{i l \varphi} \tr [\Psi^a \Omega^l \Psi^b \Omega^{J-l}]
.}
They contain the above $\UU_J^a$ and $\OO_J^a$ as ``building blocks'' and under
the ``dilute gas'' approximation the latter are the dominant pieces in the
calculation of the anomalous dimensions.

In the presence of the deformations the gauge theory equations of motion become
\eqn\deformedeom{\eqalign{
& \frac{1}{4}\bar{D}^2 \bar{\Psi}^1 = g'  [\Psi^2, \Omega] + h_1 \{ \Psi^2, \Omega \} +3h_2 (\Psi^1)^2,
\cr
& \frac{1}{4}\bar{D}^2 \bar{\Psi}^2 = - g'  [\Psi^1, \Omega] + h_1 \{ \Psi^1, \Omega \} +3h_2 (\Psi^2)^2.
}}
Notice that the gauge theory action has been expressed in terms of the
rotated basis of superfields $(\Omega,\Psi^1,\Psi^2)$. This is not necessary, but we do it here in order to comply with
the conventions adopted in section 2.3.
In the large $J$ limit the above equations imply
\eqn\deformeduoeom{\eqalign{
& \frac{1}{4}\bar{D}^2 \UU_J^1 = (g' (1-e^{-i\varphi})+h_1(1+e^{-i\varphi}))\OO^2_{J+1}+3h_2 \OO^{11}_J,
\cr
& \frac{1}{4}\bar{D}^2 \UU_J^2 = (- g' (1-e^{-i\varphi})+h_1(1+e^{-i\varphi}))\OO^1_{J+1}+3h_2 \OO^{22}_J,
}}
where we have denoted
\eqn\oaa{
\OO^{aa}_J = \sum_l e^{il \varphi} \Omega^l (\Psi^a)^2 \Omega^{J-l},
}
for $a=1,2$.
An immediate consequence is the following relation
\eqn\dudu{\eqalign{
\frac{1}{16}\langle \bar{D}^2 \UU_J^1 (z)  D^2 \bar{\UU}_J^1 (z') \rangle &=
|A_1|^2 \langle \OO^2_{J+1}(z) \bar{\OO}^2_{J+1} (z') \rangle + 9 |h_2|^2 \langle \OO^{11}_J(z) \bar{\OO}^{11}_J (z')
\rangle +
\cr
&+ 3 A_1 \bar{h}_2 \langle \OO^2_{J+1} (z) \bar{\OO}^{11}_J (z') \rangle +
3 \bar{A}_1 h_2 \langle \OO^{11}_J (z) \bar{\OO}^2_{J+1} (z') \rangle
.}}
We have defined
\eqn\Aone{
A_1=g' (1- e^{-i \varphi} ) + h_1 (1+e^{-i \varphi})
}
and $z=(x,\theta,\tbar)$ is a superspace variable.
There is a similar expression for the two-point function
$\langle \bar{D}^2 \UU_J^2 (z)  D^2 \bar{\UU}_J^2 (z') \rangle$ with
the factor $A_1$ replaced by
\eqn\Atwo{
A_2=-g' (1- e^{-i \varphi} ) + h_1 (1+e^{-i \varphi})
.}

The aim is to write down an explicit expression for each side of equation \dudu\ and use it to deduce a constraint on the
anomalous dimensions of interest.
Each side can be written down explicitly for any value of the
couplings $g, g', h_1$ and $h_2$, as long as these couplings obey the constraint \constraint\
and as long as every operator that appears in \deformeduoeom\ is quasi-primary.
We will assume that the second condition is valid throughout our calculation
even though we have not been able to find an explicit proof. Note that
the same assumption was also made in the  ${\cal N}=4$ case
\refs{\bmn, \italian}. This is a crucial ingredient
of this approach and it would be worthwhile to investigate it further. We expect it to be valid in the infinite
$J$ limit on the basis of the gauge theory/pp-wave string correspondence described in section 2.3.

Now, the key point of the computation is
that superconformal invariance determines the form of two-point functions of quasi-primary operators uniquely (up to a
normalization-dependent factor). To see how
the two-point functions look like we start with a few simple expressions at tree-level.
For a generic chiral superfield $\Phi$ we have
\eqn\twopointchiral{
\langle \Phi (z) \bar{\Phi} (z') \rangle = \frac{1}{16} \frac{1}{4\pi^2} {\bar D}^2 D^2  \frac{\delta^4(\theta-\theta')}
{|x-x'|^2}.
}
Using Wick's theorem we can further show that
\eqn\twopointfree{\eqalign{
\langle & \OO_{(h,{\bar h})}(z) \bar{\OO}_{(h,\bar{h})} (z') \rangle_{\rm free}=
\langle (\Phi^h \bar{\Phi}^{\bar h} )(z) (\bar{\Phi}^h \Phi^{\bar h} )(z') \rangle_{\rm free} =
\cr
& = c_{\OO} \bigg ( \frac{1}{16}{\bar D}^2 D^2 + \frac{ \Delta - \omega}{4\Delta} {\bar D}^{\dot {\alpha}}
\sigma_{\alpha \dot{\alpha}}^{\mu} D^{\alpha} \partial_{\mu} +
\frac {(\Delta-\omega)(\Delta-\omega-2)}{4 \Delta (\Delta-1)} \boxy \bigg )
\frac{\delta^4(\theta-\theta')}{|x-x'|^{2\Delta}}
,}}
where we have set $\Delta=h+\bar h$ for the total dimension and $\omega=h-\bar h$ for the chiral weight.
$c_{\OO}$ is an appropriate tree-level factor.

Because of superconformal invariance the same form is also valid in the interacting theory.
Given the assumption\foot{As we mentioned earlier,
this assumption appears to be valid only in the infinite $J$ limit. The BMN operators are not quasi-primary
at finite $J$ and they should receive $1/J$ corrections.}
that the operators of interest are quasi-primary in the BMN limit
the only difference between the free and the interacting cases lies in the
scaling dimensions, which may become anomalous. Similar moduli dependent contributions to
the chiral weights do not appear for the following reason.
In the presence of  $\NN=1$ superconformal invariance
the chiral weights of quasi-primary operators are proportional to their
$U(1)_R$ charges \osborn\ and the latter are not expected to receive corrections
at any order in perturbation theory.
Indeed, since the R-charge of a generic composite operator is the sum of the R-charges of its constituents (see for instance
the recent discussion in \IntriligatorJJ), R-charge corrections to the BMN operators $\UU^a_J$ and $\OO^a_J$ in \opu, \opo\
would imply that the R-charges of the constituent chiral superfields
$\Omega$, $\Psi^a$ get renormalized. This type of renormalization
cannot occur, however, because the presence of the exact $U(1)_R$ symmetry fixes the R-charge of
the perturbing superpotential \twopotential\ to be two.
The R-symmetry is part of the superconformal algebra and remains exact at any point of the moduli space.
In summary, we conclude that the only modification of \twopointfree\ in the interacting theory
is the substitution of the canonical dimension $\Delta$ by
$\Delta+\gamma$. The overall factor $c_{\OO}$ becomes in the planar limit a
function that generically depends on the
couplings $g^2N, g'^2N, h_1\sqrt{N}, h_2\sqrt{N}$ and the conformal weights $h, {\bar h}$.

Hence, the full interacting counterpart of eq. \twopointfree\
reads
\eqn\twopointint{\eqalign{
\langle & \OO_{(h,{\bar h})}(z) \bar{\OO}_{(h,\bar{h})} (z') \rangle=
c_{\OO}(g, g', h_1,h_2;N, h,\bar{h}) \bigg ( \frac{1}{16} {\bar D}^2 D^2 +
\cr
& + \frac{ \Delta +\gamma - \omega}{4(\Delta+\gamma)} {\bar D}^{\dot {\alpha}}
\sigma_{\alpha \dot{\alpha}}^{\mu} D^{\alpha} \partial_{\mu} +
\frac {(\Delta+\gamma-\omega)(\Delta+\gamma-\omega-2)}{4 (\Delta+\gamma) (\Delta+\gamma-1)} \boxy \bigg )
\frac{\delta^4(\theta-\theta')}{|x-x'|^{2(\Delta+\gamma)}}
.}}
We are now ready to
apply this general expression on the two-point functions that appear in eq.\ \dudu.

As a more straightforward situation we would like to begin with the analysis of
the special case of zero $h_2$ coupling. We will return to the more generic situation in a moment.
For $h_2=0$ eq.\ \dudu\ becomes
\eqn\sdudu{
\frac{1}{16}\langle \bar{D}^2 \UU_J^1 (z)  D^2 \bar{\UU}_J^1 (z') \rangle =
|A_1|^2 \langle \OO^2_{J+1}(z) \bar{\OO}^2_{J+1} (z') \rangle
.}
The operator $\UU_J^1$ has
canonical dimension $\Delta=J+1$, chiral weight $\omega=J-1$ and some anomalous dimension that we denote
by $\gamma_{\UU}$. As a result, we write
\eqn\duduleft{\eqalign{
\langle & \bar{D}^2 \UU_J^1 (z) D^2 \bar{\UU}_J^1 (z') \rangle = \frac{N^{J+1}}{(4\pi^2)^{J+1}}
c(g,g',h_1,h_2;N,J) \bar{D}^2 \bigg ( \frac{1}{16} \bar{D}^2 D^2 +
\cr
& + \frac{ 2 + \gamma_{\UU}}{4 (J+1+\gamma_{\UU})} {\bar D}^{\dot {\alpha}}
\sigma_{\alpha \dot{\alpha}}^{\mu} D^{\alpha} \partial_{\mu} +
\frac{(2+\gamma_{\UU})\gamma_{\UU}}{4(J+1+\gamma_{\UU})(J+\gamma_{\UU})} \boxy
\bigg ) D^2 \frac{\delta^4(\theta-\theta')}{|x-x'|^{2(J+1+\gamma_{\UU})}}=
\cr
& = \frac{N^{J+1}}{(4\pi^2)^{J+1}} c(g,g',h_1,h_2;N,J) \gamma_{\UU}(\gamma_{\UU}+2)\bar{D}^2 D^2
\frac{\delta^4(\theta-\theta')}{|x-x'|^{2(J+2+\gamma_{\UU})}}
.}}
To get the last equality we used the well-known identity $\bar{D}^3=0$.

Similarly,
the operator $\OO_{J+1}^2$ has $\Delta=\omega = J+2$ and we denote the corresponding anomalous dimension by
$\gamma_{\OO}$. In the large $J$ limit the anomalous dimensions of the operators $\OO_J^a$ and $\OO_{J+1}^a$
can be taken to be the same.
Thus,
\eqn\dudurighto{\eqalign{
\langle & \OO_{J+1}^2 (z) \bar{\OO}_{J+1}^2 (z') \rangle = \frac{N^{J+2}}{(4\pi^2)^{J+2}}
c_2(g,g',h_1,h_2;N,J) \bigg ( \frac{1}{16} \bar{D}^2 D^2 +
\cr
&+ \frac{ \gamma_{\OO}}{4 (J+2+\gamma_{\OO})} {\bar D}^{\dot {\alpha}}
\sigma_{\alpha \dot{\alpha}}^{\mu} D^{\alpha} \partial_{\mu} +
\frac{\gamma_{\OO} (\gamma_{\OO}-2)}{4 (J+2+\gamma_{\OO})(J+1+\gamma_{\OO})} \boxy
\bigg ) \frac{\delta^4(\theta-\theta')}{|x-x'|^{2(J+2+\gamma_{\OO})}} \sim
\cr
& \sim
\frac{1}{16} \frac{N^{J+2}}{(4\pi^2)^{J+2}} c_2(g,g',h_1,h_2;N,J) \bar{D}^2 D^2
\frac{\delta^4(\theta-\theta')}{|x-x'|^{2(J+2+\gamma_{\OO})}}.
}}
The last two terms have been dropped in the final equality because they are
subleading in the large $J$ limit.

When $h_2 = 0$ the normalization factors $c$ and $c_2$ are equal,
because the operators $\UU^1_{J}$ and $\OO^2_{J+1}$ are part of the same supermultiplet.
A similar situation also occurs in the $\NN=4$ and $\NN=2$ examples of refs.\ \italian, \oz.
As a result of eqs. \sdudu, \duduleft\ and \dudurighto\ we obtain the following interesting relations. First,
the simple requirement that the same power of $|x-x'|$ appears on both sides of eq.\ \dudu\ implies that
the anomalous dimensions $\gamma_{\UU}$ and $\gamma_{\OO}$ have to be the same.
Secondly, if we denote the common value of these dimensions by $\gamma^1$
we find that it has to obey the equation
\eqn\sduduresult{
(\gamma^1)^2 + 2 \gamma^1 = \frac{N}{4\pi^2} |A_1|^2
.}
$A_1$ is the constant that appears in \Aone\ and depends on $g'$, $h_1$ and $J$.
We can solve this simple quadratic equation for $\gamma^1$ and obtain an exact expression for the scaling dimension
of the operators $\UU^1_J$ and $\OO^2_J$.

Before doing that however, let us return to the
general situation of the deforming superpotential \twopotential, with both
$h_1$ and $h_2$ non-zero, and explain
what happens there.
On the r.h.s.\ of \dudu\ we have some extra two-point functions that involve the operator $\OO^{11}_J$. As we said above,
we make the explicit assumption that this operator is quasi-primary in the infinite $J$ approximation. This allows
the use of the general equation \twopointfree. For the operator $\OO^{11}_J$ we have
$\Delta=\omega=J+2$. We denote its anomalous scaling dimension by $\gamma_{\OO^{11}}$ and,
similar to the two-point function of $\OO_J^2$, we get
\eqn\dudurightoo{
\langle  \OO_J^{11} (z) \bar{\OO}_J^{11} (z') \rangle \sim
\frac{1}{16} \frac{N^{J+2}}{(4\pi^2)^{J+2}}
c_{11}(g,g',h_1,h_2;N,J) \bar{D}^2 D^2
\frac{\delta^4(\theta-\theta')}{|x-x'|^{2(J+2+\gamma_{\OO^{11}})}}.
}
Moreover, we will allow for a non-diagonal overlap between the quasi-primary
operators $\OO_{J+1}^2$ and $\OO_J^{11}$ by setting
\foot{In perturbation theory these operators
have equal scaling dimensions at weak coupling and
we have to consider the possibility of mixing.
Indeed, in higher orders of perturbation theory
this seems possible. We will not discuss this perturbative mixing in this section.
Instead, we make use of the large $J$ assumption that
the operators in question are quasi-primary and on general grounds proceed with the application
of eq.\ \twopointfree.}
\eqn\mixing{
\langle  \OO_{J+1}^2 (z) \bar{\OO}_J^{11} (z') \rangle \sim
\frac{1}{16} \frac{N^{J+2}}{(4\pi^2)^{J+2}} c_{12}(g,g',h_1,h_2;N,J) \bar{D}^2 D^2
\frac{\delta^4(\theta-\theta')}{|x-x'|^{2(J+2+\frac{1}{2}(\gamma_{\OO^{11}}+\gamma_{\OO}))}}
,}
where $c_{12}$ is a certain function of the couplings.

In this more generic situation, the operator $\UU^1_J$
is part of the same supermultiplet as the linear combination of the operators that appear on the r.h.s. of
the equations of motion \deformeduoeom.
This statement alone, however, is not enough to determine the relation between the
normalization factors $c,c_2,c_{11}$ and $c_{12}$.
For the special case $h_2 = 0$ the operator $\OO^{11}_{J}$ was absent and we could deduce the relation $c=c_2$.
In addition, the leading order perturbative computation of the next
subsection yields
$c=c_2=c_{11}$ also with $h_2 \neq 0$. This indicates that the
equation $c=c_2=c_{11}$ is true at a
generic point of the moduli space.
The present approach, however, does not provide a proof of this fact and
in this subsection we are forced to work with generic normalization factors.

Combining the information of
eqs.\ \dudu, \duduleft, \dudurighto, \dudurightoo, and \mixing\ we find that
the anomalous dimensions $\gamma_{\UU},\gamma_{\OO}$ and $\gamma_{\OO^{11}}$ have to be the same and we
denote them by $\gamma^1$. Again, this result has been derived in the large $J$ limit and we expect
$1/J$ corrections to lift this degeneracy.
Furthermore, we find the generalization of eq.\ \sduduresult\
\eqn\duduresult{
(\gamma^1)^2 + 2 \gamma^1 = \frac{N}{4\pi^2} \big(\frac{c_2}{c} |A_1|^2+9 \frac{c_{11}}{c} |h_2|^2
+3A_1 \bar{h}_2 \frac{c_{12}}{c}+3\bar{A_1} h_2 \frac{\bar{c}_{12}}{c}\big)
.}
In the large 't-Hooft and large $J$ limit
the ratios of the $c$-functions are going to be
functions of the couplings $g'^2,h_1,h_2$ and $g^2$ multiplied by the appropriate power of $N$ or $J$.
We denote them as
\eqn\ratios{\eqalign{
\FF_{2}(g, g',h_1, h_2;N,J) &= \frac{c_{2}}{c} \ ,
\cr
\FF_{11}(g, g',h_1, h_2;N,J) &= \frac{c_{11}}{c} ,
\cr
\FF_{12}(g, g',h_1, h_2;N,J) &= \frac{c_{12}}{c} .
}}
They are expected to take finite values in the large $J$ limit, but, as we said, in generic points of the moduli
space it is not possible to determine them exactly using the technology of \italian.
Their leading order behavior will be determined in perturbation theory in the next subsection.

With this notation
eq.\ \duduresult\ has only one reasonable solution and we can write it as
\eqn\gammasolution{
\gamma^1=-1+\sqrt{1+\frac{N}{4\pi^2} (\FF_{2} |A_1|^2+9\FF_{11}|h_2|^2
+3A_1 \bar{h}_2 \FF_{12}+3\bar{A_1} h_2 \bar{\FF}_{12})}.
}

Before expressing $A_1$ more explicitly in the large $J$ approximation, it is probably useful to emphasize
the following well-known fact. In the
large $N$ limit it is natural to scale the deformation couplings $h_1$ and
$h_2$ as $1/\sqrt{N}$. Indeed, as was pointed out, for example
in \intriligator, in the large 't Hooft limit it is convenient to normalize
the $\NN=4$
chiral primary operators $\OO_p$
with Dynkin labels $(0,p,0)$
as
\eqn\deformscaling{
\OO_p=N(g_{YM}^2 N)^{-p/2} \tr [\Phi^{(i_1} \cdot \cdot \cdot \Phi^{i_p)} ]
.}
A similar normalization for the $\NN=1$ chiral primary operators
that appear in \twopotential, gives
the extra factor $\frac{1}{g_{YM}^3 \sqrt{N}}$ and since
we keep $g_{YM}$ small and fixed we simply set $h_i =
\frac{\lambda_i}{\sqrt{N}}$ for $i=1,2$.
The scaling of the
overall coefficient $g'$ in front of the $\NN = 4$ superpotential is a bit
more subtle. In the large $J$ and large 't-Hooft limit it is natural to write $g'=g+h_0$, with
$g=\sqrt{2} g_{YM} << 1$ being the $\NN=4$ value of the coupling and treat
$h_0$ on the same footing as $h_1$ and $h_2$. Hence, we also scale $h_0$ as $1/\sqrt{N}$ and we set
$h_0=\lambda_0/\sqrt{N}$.
The reasons for this choice will become more apparent in the next subsection, where we discuss the perturbative form
of the constraint \constraint. As a consequence of these scalings,
$g'$ can be simply substituted by $g$ in any expression, where the dominant $g$ dependence does not cancel exactly.

With these conventions we can write the large $N$ and $J$ limit of $\gamma^1$ as
\eqn\fullgammasolutionone{
\gamma^1=-1+\sqrt{1+\alpha_1^2-\alpha_2
\frac{g\sqrt{\FF_2}\sqrt{N}n}{J} + \frac{\FF_2 g^2 N n^2}{J^2}}
,}
where we have set
\eqn\alphas{\eqalign{
& \alpha_1^2 = \frac{1}{4\pi^2}(4\FF_2|\lambda_1|^2+9\FF_{11}|\lambda_2|^2+6\FF_{12} \lambda_1 \bar{\lambda}_2
+6 \bar{\FF}_{12} \bar{\lambda}_1 \lambda_2),
\cr
& \alpha_2 =  - \frac{2}{\pi \sqrt{\FF_2}} ({\rm Im}(\lambda_1)
+\frac{3}{4}i( \FF_{12} \bar{\lambda}_2 - \bar{\FF}_{12} \lambda_2))
.}}
The functions $\FF_2,\FF_{11},\FF_{12}$ can in principle depend
only on the couplings $\lambda_0, \lambda_1, \lambda_2$ and the finite ratio
$g^2 N/J^2$.  Use of the constraint \constraint\ should allow a further elimination of the dependence on one of these couplings.

In precisely the same way one may also calculate the anomalous dimension $\gamma^2$ of the operators $\UU_J^2$ and
$\OO_J^1$. The result is
\eqn\fullgammasolutiontwo{
\gamma^2=-1+\sqrt{1+\alpha_1^2+\alpha_2
\frac{g\sqrt{\FF_2}\sqrt{N}n}{J} + \frac{\FF_2 g^2 N n^2}{J^2}}
.}
This dimension is different from $\gamma^1$.
This is another effect of the breaking of the $\NN=4$ superconformal symmetry down
to $\NN=1$. The corresponding anomalous dimensions at the $\NN=4$ point are the same as a consequence of the
extended supersymmetry. In particular, it has been shown in \beisert, that
all the relevant BMN operators with two $\Psi^a$ insertions belong to the same long supermultiplet.
After the $\NN=4$-breaking deformation this property disappears. As we see later,
from the string theory perspective this difference is due to the
appearance of 3-form fluxes along the $(\Psi^1,\Psi^2)$ plane
which break the transverse $SO(4)$ symmetry.

Finally, notice that for generic points of the moduli space the anomalous dimensions \fullgammasolutionone\ and
\fullgammasolutiontwo\ may become imaginary. Using the perturbative values of $\FF_2,\FF_{11}$ and $\FF_{12}$ we
can see that this is not happening around the $\NN=4$ point. If it
does happen deeper into the moduli space
it is not necessarily a bad or pathological feature of the theory, but it should be rather
interpreted as a sign that we have to use a different coordinate system on the field theory space
in order to get a reasonable description.

\subsec{Two-point functions and scaling dimensions in perturbation theory}

In this subsection we make an independent computation of
the anomalous scaling dimensions \fullgammasolutionone\ and
\fullgammasolutiontwo\ to leading order in perturbation
theory. This will also provide the perturbative values of the ratios \ratios.

We work in superspace formalism and consider only planar diagrams.
Our perturbative treatment involves four parameters.
Two of them are related to the
marginal deformations parameterized by $h_1$ and $h_2$. In the previous
subsection we re-expressed them as $\lambda_1$ and $\lambda_2$ and they are
finite and tunable couplings in the planar limit. There are two more. One of them we denoted by
$g'=g+h_0$ and we set $h_0=\lambda_0/\sqrt{N}$ and the other
parameter is proportional to the Yang-Mills coupling.
By now, it has been firmly established
\refs{\staudacher,\motl}
that what governs the strong t'-Hooft, large R-charge perturbation theory with respect to the $\NN=4$
superpotential
is not t' Hooft's coupling per se, which becomes infinite, but rather
the finite and tunable parameter
$\lambda' = g^2_{YM} \frac{N}{J^2}$.

We begin by computing the leading order correction to the two-point function
$\langle \UU_J^1 (z) \;\bUU_J^1 (z') \rangle$
in the presence of the exactly
marginal deforming superpotential  \twopotential\
\eqn\twopotential{W_{def} =
h_1 {\rm Tr} (\Psi^1 \Psi^2 \Omega + \Psi^2 \Psi^1 \Omega) +
h_2 {\rm Tr} \big({(\Psi^1)}^3+{(\Psi^2)}^3+ \Omega^3 \big). }
The interacting part of the full action,
which involves only the Higgs superfields $\Psi^1, \Psi^2, \Omega$ can be written as
\eqn\deformaction{\eqalign{
\int d^4 x \int d^2 \theta \Big[ \;g' {\rm Tr} & \big( (\Psi^1 \Psi^2 - \Psi^2
\Psi^1)  \Omega \big) +
\cr
& h_1 {\rm Tr}  \big( (\Psi^1 \Psi^2 + \Psi^2 \Psi^1)  \Omega \big) +
  h_2 {\rm Tr} \big({(\Psi^1)}^3+{(\Psi^2)}^3+ \Omega^3 \big) \Big] + {\rm
c.c.}
}}
and the leading non-zero corrections come from the second order terms\foot{
We use the convention that the
superspace coordinate $z=(x,\theta,\tbar)$ is appropriately
truncated to its chiral part $(x,\theta)$
when it appears as an
argument of a chiral superfield and similarly for antichiral superfields.}
\eqn\leadhone{\eqalign{
&  \int d^4 x_1 \int d^2 \theta_1  \int d^4 x_2 \int d^2 \tbar_2
\;\;
\Big[    (h_1 +g')({\bar h_1}-\bar {g'})
{\rm Tr}(\Psi^1 \Psi^2 \Omega ( z_1 )) \;
{\rm Tr}(\bPsi^1 \bPsi^2 \bO ( z_2 )) +
\cr
&
 |h_1 +g'|^2
{\rm Tr}(\Psi^1 \Psi^2 \Omega ( z_1 )) \;
{\rm Tr}(\bPsi^2 \bPsi^1 \bO ( z_2 )) +
     |h_1-g'|^2
{\rm Tr}(\Psi^2 \Psi^1 \Omega ( z_1 )) \;
{\rm Tr}(\bPsi^1 \bPsi^2 \bO ( z_2 )) +
\cr
&
(h_1-g')({\bar h_1}+\bar {g'})
{\rm Tr}(\Psi^2 \Psi^1 \Omega ( z_1 )) \;
{\rm Tr}(\bPsi^2  \bPsi^1 \bO ( z_2 ))
\Big]
}}
and
\eqn\leadhtwo{
 |h_2|^2
\int d^4 x_1 \int d^2 \theta_1  \int d^4 x_2 \int d^2 \tbar_2
\;\; \Big[ {\rm Tr}{(\Psi^1)}^3 (z_1)  {\rm Tr}{(\bPsi^1)}^3 (z_2)
+  {\rm Tr}\Omega^3 (z_1)  {\rm Tr}\bO^3 (z_2) \Big) \Big].
}
Note that there is no mixing between the $h_1$ and $h_2$
deformations at this order in perturbation theory.
The contribution
to the anomalous dimensions from diagrams involing the gauge field
multiplet will be taken into account at the end of the computation.

The general form \opu\ of the operators $\UU_J^1$ implies that we need
to compute amplitudes of the form $\langle \Omega^l \bPsi^1 \Omega^{J-l} (z)
\; \bO^{m} \Psi^1 \bO^{J-m} (z') \rangle$ with insertions of
the interacting terms \leadhone\ and \leadhtwo.
In order to find the full two-point function
$\langle \UU_J^1 (z) \;\bUU_J^1 (z') \rangle$ we have to incorporate the
phases $e^{i (l-m) \varphi}$
and sum over the integers $l$ and $m$ that provide planar diagrams.
In order to make the computation more transparent
we first ignore the fact that the fields are matrices in
the $SU(N)$ Lie algebra and reinstate the relevant group-theoretical
factors later.

After using Wick's theorem and the chiral superfield propagators \twopointchiral,
we obtain three types of diagrams
from \leadhone. We can write them as
\eqn\diagram{\eqalign{
\bullet \;&\; \frac{1}{16^{J+4}}\frac{1}{(4 \pi^2)^{J+4}}
\int d^4 x_1 \int d^2 \theta_1  \int d^4 x_2 \int d^2 \tbar_2
\big(\bD^2 D^2 F(z,z') \big)^{J-1}
\big(D^2 \bD^2 F(z,z') \big)
\cr
&
\big(D^2 \bD^2 F(z_2,z) \big)
\big(\bD^2 D^2 F(z_1,z') \big)
\big(\bD^2 D^2 F(z_1,z_2) \big)^2 \cr
\bullet \;&\; \frac{1}{16^{J+4}}\frac{1}{(4 \pi^2)^{J+4}}
\int d^4 x_1 \int d^2 \theta_1  \int d^4 x_2 \int d^2 \tbar_2
\big(\bD^2 D^2 F(z,z') \big)^J
\big(\bD^2 D^2 F(z_1,z) \big)
\cr
&
\big(D^2 \bD^2 F(z_2,z') \big)
\big(\bD^2 D^2 F(z_1,z_2) \big)^2
\cr
\bullet \;&\; \frac{1}{16^{J+4}}\frac{1}{(4 \pi^2)^{J+4}}
\int d^4 x_1 \int d^2 \theta_1  \int d^4 x_2 \int d^2 \tbar_2
\big(\bD^2 D^2 F(z,z') \big)^{J-1}
\big(D^2 \bD^2 F(z_2,z) \big) \cr
&
\big(\bD^2 D^2 F(z_1,z') \big)
\big(\bD^2 D^2 F(z_1,z_2) \big)
\big(D^2 \bD^2 F(z_2,z') \big)
\big(\bD^2 D^2 F(z_1,z) \big) .
}}

For convenience we use the notation
\eqn\conveprop{
F(z,z') = \frac{\delta^4(\theta-\theta')} {|x-x'|^2}
}
and by convention the superspace derivatives $\bD$ and $D$ are taken to act
on the first argument of $F$. The associated super-Feynman
diagrams are shown in Figure 1.
The first two diagrams are wave-function
renormalizations of the fields $\Omega$  and $\Psi^1$
respectively and they are phase-independent in the planar
limit.
The third diagram
interchanges the position of $\bPsi^1$ and is responsible for the level $n$
dependence of the anomalous dimension.

\vskip 0.4 cm
\multiply \epsfxsize by 2
\centerline{\epsfbox{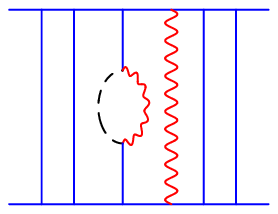} \epsfbox{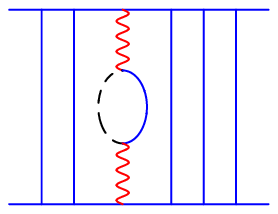}
\epsfbox{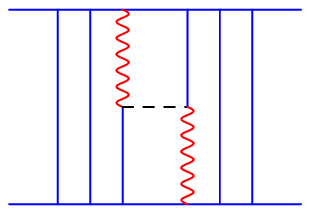}}
\vskip 0.4 cm
{\bf Fig. 1.} {\it The three types of super-Feynman diagrams that contribute to the leading order anomalous dimensions
of $\UU_J^1$.
Straight lines depict propagators of $\Omega$, wiggly lines propagators of $\Psi^1$ and syncopated lines propagators
of $\Psi^2$.}
\vskip 0.4 cm

In order to compute the anomalous dimension of an operator to leading order we
are instructed to compute the relevant two-point function and, if working
in dimensional regularization, extract the $1/\epsilon$ divergence.
In a different regularization scheme, e.g. with a UV cutoff,
this divergence corresponds to a logarithmic correction of the propagator.
After the appropriate renormalization, the final result takes the generic form
\twopointint.

A well-known subtlety in the above procedure has to do with
operators that share the same canonical dimensions.
Usually such operators mix on the quantum level
and the corresponding two-point functions are no longer diagonal.
In that case, the correct anomalous dimensions
result from the diagonalization of the mixing matrix.
This appears to be a problem in our case because it seems that in general we have to diagonalize an infinite dimensional
mixing matrix. The same problem is also encountered at the $\NN=4$ point. Our attitude towards this is
the following. As in section 3.3 we make the large $J$ assumption
that the operators in question are always quasi-primary and this allows the
computation of the anomalous dimensions from a
simple two-point function calculation.

In our superspace formalism computation it is convenient
to focus on the theta-independent piece of the superfield
two-point function.
This corresponds to the two-point function between the
bottom components of each operator and by supersymmetry all the
fields in the full multiplet should have the same anomalous dimensions.
In general, if the 1-loop
corrected propagator between the bottom components
of a superfield operator $\OO_{(h,\bar{h})}$ with canonical dimension
$\Delta=h+{\bar h}$
has the form
\eqn\correctedprop{
\langle \OO_{(h,{\bar h})}|_{\theta=\tbar=0}(z) \;
\bar{\OO}_{(h,\bar{h})}|_{\theta'=\tbar'=0} (z') \rangle=
c_{\OO}^{(1)} \frac{1}{|x-x'|^{2\Delta}}
\big(1+ \gamma_{(1)} \frac{1}{\epsilon}\big) ,
}
the leading contribution to its
anomalous dimension is given by the coefficient  $\gamma_{(1)}$
of the UV divergent term. The divergence is regularized
by the standard dimensional regularization continuation to $d=4-2\epsilon$ dimensions.

We can now proceed with the computation of the
first diagram. The free piece is given by the general
form \twopointfree\ after setting $\Delta=J$ and $\omega=J-2$ and
in the planar limit the associated factor $c_{\OO}$ is $(\frac{N}{4 \pi^2})^J$.
The computation of
the superspace integral involves the free chiral superfield propagators \wb
\eqn\freeprops{\eqalign{
\bD^2 D^2 F(z_1,z_2) &= 16 e^{i(\theta_1 \sigma^n \tbar_1 +
\theta_2 \sigma^n \tbar_2 -2 \theta_1 \sigma^n \tbar_2) \partial_n}
\frac{1}{|x_1-x_2|^2} \cr
D^2 \bD^2 F(z_1,z_2) &= 16 e^{-i(\theta_1 \sigma^n \tbar_1 +
\theta_2 \sigma^n \tbar_2 -2 \theta_2 \sigma^n \tbar_1) \partial_n}
\frac{1}{|x_1-x_2|^2} \cr
}.}
Remember that we
always take the superspace derivatives to act on the
first argument of $F(z_1,z_2)$.

Then we can easily perform the fermionic integrations to obtain
\eqn\firstinthone{\eqalign{
&
 \int d^2 \theta_1  \int d^2 \tbar_2 \;\;
\big(D^2 \bD^2 F(z_2,z) \big)
\big(\bD^2 D^2 F(z_1,z') \big)
\big(\bD^2 D^2 F(z_1,z_2) \big)^2 = \cr
&
16^4 \big[\boxy \D_{x_1 x_2}^2
\big(\D_{x x_2} -i \theta \sigma^n \tbar \partial_n^{x_2} \D_{x x_2}
+\frac{1}{4} \theta^2 \tbar^2 \boxy  \D_{x x_2}\big)
\big(\D_{x' x_1} -i \theta' \sigma^m \tbar' \partial_m^{x_1} \D_{x' x_1} \cr
&
+\frac{1}{4} \theta'^2 \tbar'^2 \boxy  \D_{x' x_1}\big)
\big] + \big[\theta^2 \tbar^2 \D_{x_1 x_2}^2
\boxy \D_{x x_2}  \boxy \D_{x' x_1}\big]
+ \big[4 i \D_{x_1 x_2} \big(i \theta \partial_n^{x_2} \D_{x x_2} +
\cr
&+(\theta \sigma^l \tbar) \theta \partial_l^{x_2} \partial_n^{x_2} \D_{x x_2}
\big)
\sigma^n \partial_m^{x_2} \D_{x_1 x_2} {\bar \sigma}^m \sigma^k
\big(i \tbar' \partial_k^{x_1} \D_{x' x_1} + (\theta' \sigma^s \tbar')
\tbar' \partial_s^{x_1} \partial_k^{x_1}\big)\big] ,
}}
where
\eqn\shortprop{
\D_{x_1 x_2} = \frac{1}{|x_1-x_2|^2} .
}

The spacetime integrations of the theta-independent piece give
\eqn\tindonehone{
\int d^4 x_1  \int d^4 x_2 \; 16^4
\D_{x x_2} \D_{x' x_1} \boxy \D_{x_1 x_2}^2 = 16^4  \frac{8 \pi^4}{\epsilon}
\frac{1}{|x-x'|^2}
}
where we have used
dimensional regularization and the general formula (see, for example, \petkou )
\eqn\dimreg{
\int d^d x \frac{1}{|x|^{2a} |x-y|^{2b}} = \pi^2
\frac{\Gamma(a+b-\frac{d}{2})}{\Gamma(a) \Gamma(b)}
\frac{\Gamma(\frac{d}{2}-a) \Gamma(\frac{d}{2}-b)}{\Gamma(d-a-b)}
\frac{1}{|y|^{2(a+b-\frac{d}{2})}} ,
}
in $d=4-2\epsilon$ dimensions.

The second diagram can be performed in a similar way. Basically,
the integrand differs from the one we just computed by the
interchange $z \leftrightarrow z'$ and the theta-independent piece gives again
\tindonehone\ . The contribution of the free contractions is
of the type \twopointfree\ with $\Delta=\omega=J$ and
the relevant prefactor in the planar limit is  $(\frac{N}{4 \pi^2})^J$.
It is also important to notice that for both the first and second diagram there
is only one planar contraction, which comes from the second
and third term in \leadhone.
In the two-point function  $\langle \Omega^l \bPsi^1 \Omega^{J-l} (z)
\; \bO^{m} \Psi^1 \bO^{J-m} (z') \rangle$ this requires $l=m$.
As a result, such diagrams do not depend on the level $n$ and the summation over
$l$ and $m$ just gives a factor of $J$ which is common
to all diagrams and which along with the same factor of the tree
level correlator can be absorbed in the overall normalization
of the operator \bmn\ .

Because all the fields under consideration are in the adjoint
representation of the $SU(N)$ gauge group, we also have to take into
account the relevant group theory factors. For a field
$\Phi_{a b} = \Phi_{A} T^{A}_{a b}$, with $A=1,\ldots,N^2-1$
and $T^{A}_{a b}$ a set of $N \times N$ traceless and hermitian
matrices spanning the Lie algebra of $SU(N)$, the free propagator
comes with a factor
\eqn\grouptheory{
\langle \Phi_{a b} {\bar \Phi}_{a' b'} \rangle \sim
(\delta_{a b'} \delta_{a' b} - \frac{1}{N}
\delta_{a b} \delta_{a' b'}).
}
The second term at the right is subleading in large $N$ computations
and can be neglected. Due to summation over indices in the loop,
contractions involving the interaction vertices come with
an additional factor of $N$ compared
to the free propagator,
and after incorporating them into the full
amplitude they result to an extra factor of $N^2$.

Putting everything together, we conclude
that the UV divergent term in the two-point function of the bottom components coming from the first
and the second diagram is
\eqn\totaloneplustwohone{\eqalign{
(1+1) \;2 (|h_1|^2+|g'|^2) \;
\Big(\frac{N}{4 \pi^2}\Big)^J \;\frac{1}{(4 \pi^2)^4} \;N^2
\;\frac{8 \pi^4}{\epsilon} \;\frac{1}{|x-x'|^{2(J+1)}} = \cr
\big(\frac{N}{4 \pi^2}\big)^{J+1} \;\frac{1}{\epsilon}\;
\Big[\frac{N}{2 \pi^2} \;(|h_1|^2+|g'|^2) \Big]\;\frac{1}{|x-x'|^{2(J+1)}}.
}}

The contribution of the third diagram can be deduced in a similar way.
The free contractions in this case take the form
\twopointfree\ with $\Delta=\omega=J-1$ and there is a usual
prefactor
 $(\frac{N}{4 \pi^2})^{J-1}$. The loop integral is
now more complicated and we write down explicitly
only the theta-independent part, which is sufficient
for our purposes. This reads
\eqn\tindtwohone{
\int d^4 x_1  \int d^4 x_2 \; 16^5 \D_{x x_1} \D_{x' x_1} \D_{x x_2}
\D_{x' x_2} \boxy \D_{x_1 x_2} =   16^5  \frac{8 \pi^4}{\epsilon}
\frac{1}{|x-x'|^4}
.}
We used the identity $\boxy \D_{x_1 x_2} = - 4 \pi^2 \delta(x_1 - x_2)$
and the general formula \dimreg\ .

Since we work in the planar limit,
only amplitudes with  $l-m=1$ from the
first vertex  and
$l-m=-1$ from the fourth vertex  of  \leadhone\
should be kept. Consequently, this diagram
is phase-dependent and the relevant factor is
$e^{i \varphi} (h_1+g')({\bar h_1}-\bar{g'})
+ e^{-i \varphi}  (h_1-g')({\bar h_1}+\bar{g'}) =
2 \cos\varphi \;(|h_1|^2 - |g'|^2) + 4 \sin\varphi {\rm {\rm Im}}(\bar{g'} h_1).$
The total group theoretical factor
is $N^3$.

Assembling every piece from the third diagram
we get the following contribution to the bottom component UV divergence
\eqn\totalthreehone{\eqalign{
\big(
2 \cos\varphi \;(|h_1|^2 - |g'|^2) + 4 \sin\varphi \; {\rm {\rm Im}}(\bar{g'} h_1)
\big)
\;
\big(\frac{N}{4 \pi^2}\big)^{J-1} \;\frac{1}{(4 \pi^2)^5} \;N^3
\;\frac{8 \pi^4}{\epsilon} \;\frac{1}{|x-x'|^{2(J+1)}} = \cr
\big(\frac{N}{4 \pi^2}\big)^{J+1} \;\frac{1}{\epsilon}\;
\Big[\frac{N}{4 \pi^2} \;
\big(
 \cos\varphi \;(|h_1|^2 - |g'|^2) + 2 \sin\varphi \; {\rm {\rm Im}}(\bar{g'} h_1)
\big)
\Big]
\;\frac{1}{|x-x'|^{2(J+1)}} .
}}
As a result, the total correction due to
\leadhone\ can be written as
\eqn\totalhone{
\big(\frac{N}{4 \pi^2}\big)^{J+1} \;\frac{1}{\epsilon}\;
\Big[\frac{N}{4 \pi^2} \;
\big(
(2+ \cos\varphi) \;|h_1|^2 +
(2- \cos\varphi) \; |g'|^2 +
2 \sin\varphi \; {\rm {\rm Im}}(\bar{g'} h_1)
\big)
\Big]
\;\frac{1}{|x-x'|^{2(J+1)}} .
}

A similar analysis can be performed for the diagrams that come from
\leadhtwo. From the first term we obtain only one connected
diagram,
which is identical to the second diagram from
\leadhone\ and an extra symmetry factor of 9.
In the planar limit only amplitudes with $l=m$ should be kept.
From the second term we get two diagrams;
one is exactly the same as the first diagram of \leadhone\
and
the other is given by
\eqn\diagramhtwo{\eqalign{
\bullet \;&\; \frac{1}{16^{J+4}}\frac{1}{(4 \pi^2)^{J+4}}
\int d^4 x_1 \int d^2 \theta_1  \int d^4 x_2 \int d^2 \tbar_2
\big(\bD^2 D^2 F(z,z') \big)^{J-2}
\big(D^2 \bD^2 F(z,z') \big) \cr
&
\big(\bD^2 D^2 F(z_1,z') \big)^2
\Big(D^2 \bD^2 F(z_2,z') \big)^2
\big(\bD^2 D^2 F(z_1,z_2) \big)
.}}
Again, for both of them there is an extra symmetry factor of 9.

Computing the theta-independent term we get
\eqn\tindonehtwo{
16^5 \int d^4 x_1  \int d^4 x_2 \;
\D_{x x_2}^2 \D_{x' x_1}^2 \boxy \D_{x_1 x_2} = 16^5  \frac{8 \pi^4}{\epsilon}
\frac{1}{|x-x'|^4}.
}
After combining everything, we conclude
that the divergent piece in the
bottom component of the two-point function due to \leadhtwo\ is
\eqn\totalhtwo{\eqalign{
9 |h_2|^2\; (1+1+1)
\big(\frac{N}{4 \pi^2}\big)^J \;\frac{1}{(4 \pi^2)^4} \;N^2
\;\frac{8 \pi^4}{\epsilon} \;\frac{1}{|x-x'|^{2(J+1)}} = \cr
\big(\frac{N}{4 \pi^2}\big)^{J+1} \;\frac{1}{\epsilon}\;
\Big[\frac{N}{4 \pi^2} \;
\frac{27}{2} |h_2|^2 \Big]
\;\frac{1}{|x-x'|^{2(J+1)}}.
}}

Adding \totalhone\ with  \totalhtwo\ and comparing with
the general expression \correctedprop\ gives the value of
the anomalous dimension to first order
\eqn\gammafirst{
\gamma^1_{(1)} = \frac{N}{4 \pi^2}
\Big[\Big(
(2+ \cos\varphi) \;|h_1|^2 +
(2- \cos\varphi) \; |g'|^2) +
2 \sin\varphi \; {\rm {\rm Im}}(\bar{g'} h_1)
\Big)+\Big(\frac{27}{2} |h_2|^2\Big)\Big].
}

In order to compare this expression to the exact result
\fullgammasolutionone\  of subsection 3.2 we have to take into account the
contribution of diagrams involving the gauge field multiplet,
which have been ignored so far, and we also have to use the leading order form of the constraint
\constraint\ to eliminate the $\lambda_0$ dependence and express the result
only in terms of $\lambda_1, \lambda_2$ and $g$. The vector multiplet contribution is, of course, $n$-independent and as
in \bmn\ it can be simply deduced from the requirement that at the $\NN=4$
point the operators without phases
are protected and hence they have vanishing anomalous
dimensions. This implies that we have to make a
shift proportional to the gauge coupling $g^2$
in the previous expression, so that the final result reads
\eqn\gammafirstsub{
\gamma^1_{(1)} = \frac{N}{4 \pi^2}
\Big[\Big(
(2+ \cos\varphi) \;|h_1|^2 +
(2- \cos\varphi) \; |g'|^2 - g^2 +
2 \sin\varphi \; {\rm {\rm Im}}(\bar{g'} h_1)
\Big)+\Big(\frac{27}{2} |h_2|^2\Big)\Big].
}
The leading order form of the constraint equation \constraint\ can be determined by computing
the perturbative anomalous dimension of the operator $\Pi_J$ \groundoperator\ and require that it vanishes.
We now proceed to determine this anomalous dimension.

Perturbative corrections to the two-point function
$\langle \Omega^J (z) \; \bO^J (z') \rangle$ are once again due to
\leadhone\ and \leadhtwo\ . From \leadhone\ we obtain one diagram
that corresponds to a wave-function renormalization of $\Omega$. This is
similar to the first diagram we encountered in the computation of
$\langle \UU_J^1 (z) \;\bUU_J^1 (z') \rangle$ and can be written as
\eqn\diagramoneground{\eqalign{
\bullet \;&\; \frac{1}{16^{J+2}}\frac{1}{(4 \pi^2)^{J+2}}
\int d^4 x_1 \int d^2 \theta_1  \int d^4 x_2 \int d^2 \tbar_2
\big(\bD^2 D^2 F(z,z') \big)^{J-1}
\big(D^2 \bD^2 F(z_2,z) \big) \cr
&
\big(\bD^2 D^2 F(z_1,z') \big)
\big(\bD^2 D^2 F(z_1,z_2) \big)^2 .
}}
The corresponding theta-independent piece has the UV divergent term
\eqn\groungtotalhone{\eqalign{
\;2 (|h_1|^2+|g'|^2) \;
\big(\frac{N}{4 \pi^2}\big)^{J-1} \;\frac{1}{(4 \pi^2)^4} \;N^2
\;\frac{8 \pi^4}{\epsilon} \;\frac{1}{|x-x'|^{2J}} = \cr
\big(\frac{N}{4 \pi^2}\big)^{J} \;\frac{1}{\epsilon}\;
\Big[\frac{N}{4 \pi^2} \;(|h_1|^2+|g'|^2) \Big]\;\frac{1}{|x-x'|^{2J}} .
}}

From the second term of \leadhtwo\ we obtain two diagrams.
These are similar to the corresponding diagrams in the
previous computation of $\langle \UU_J^1 (z) \;\bUU_J^1 (z') \rangle$,
except that the former have an extra propagator $\big(D^2 \bD^2 F(z,z') \big)$
due to the $\Psi^1$ insertions. Comparison with \totalhtwo\
gives the anomalous dimension contribution
\eqn\totalhtwo{\eqalign{
9 |h_2|^2\; (1+1)
\big(\frac{N}{4 \pi^2}\big)^J \;\frac{1}{(4 \pi^2)^4} \;N^2
\;\frac{8 \pi^4}{\epsilon} \;\frac{1}{|x-x'|^{2(J+1)}} = \cr
\big(\frac{N}{4 \pi^2}\big)^{J+1} \;\frac{1}{\epsilon}\;
\Big[\frac{N}{4 \pi^2} \;
\frac{18}{2} |h_2|^2 \Big]
\;\frac{1}{|x-x'|^{2(J+1)}}.
}}

Consequently, the leading order correction to the anomalous
dimension of $\Pi_J$ is
\eqn\groundgamma{
\gamma_{\Pi_{J} (1)} = \frac{N}{4\pi^2} (|g'|^2-g^2+|h_1|^2+9|h_2|^2) .
}
As before, we have subtracted an appropriate constant term in order to account
for the contribution of diagrams involving
the gauge multiplet. As we said, superconformal invariance requires the vanishing of this anomalous dimension.
The resulting constraint is the leading order form of the constraint equation \constraint\
\eqn\pertconstraint{
g'^2N-g^2N+|\lambda_1|^2+9|\lambda_2|^2=0
.}
This relation can also be found in \refs{\razamattwo, \razamat}.
We can use it to eliminate the $h_0$ dependence from the anomalous
dimension \gammafirstsub\ and thus we get
\eqn\gammafirstcomple{
\gamma^1_{(1)} = \frac{N}{4 \pi^2}
\Big[
2 \;|\lambda_1|^2 +\frac{9}{2} |\lambda_2|^2 +
\frac{1}{\pi} {\rm {\rm Im}}(\lambda_1) \frac{g\sqrt{N}n}{J}+
\frac{1}{2} \frac{g^2 N n^2}{J^2}
\Big].
}
This result agrees with the
outcome of the exact computation
\fullgammasolutionone\ of the previous subsection and also provides the leading order
values of the ratios \ratios:
\eqn\leadingF{
\FF_{2}=\FF_{11}=1 \  {\rm and} \ \FF_{12}=0
.}

\newsec{String theory backgrounds from gauge theory anomalous dimensions}

The purpose of this section is the reconstruction of a string theory from the gauge theory data
that were obtained above. The working assumption of our analysis is that the correspondence
between gauge theory operators and string states that was proposed in \bmn\ and outlined in our case
in section 2 remains valid even when one deforms away from the $\NN=4$ point. We have traced the
effect of the deformation on the gauge theory side and now we would like to trace this effect also on
the string theory side. In the large $J$ limit this is possible because of the nature of the gauge theory/string theory
correspondence, which essentially amounts to the identification of the scaling weights of certain near-BPS
operators on the gauge theory side with the spectrum of a light-cone worldsheet theory. This allows
the reverse-engineering of a ``dual'' string background directly from gauge theory data. The
word dual is inside quotation marks, because we will soon see that this reverse-engineering process
does not produce a unique background in the infinite $J$ limit.

\subsec{Bosonic sector}

The bosonic part of the light-cone worldsheet theory at the $\NN=4$ point is given by the action
\eqn\bosonicaction{
\SS_0 = \SS_{\vec{r},0}+\SS_{z^1,0}+\SS_{z^2,0}
,}
where there are three distinct parts. $\SS_{\vec{r},0}$ is related to the four directions that descend from the $AdS_5$
part of the geometry after we take the Penrose limit and reads
\eqn\srizero{
\SS_{\vec{r},0}=\frac{1}{2\pi \alpha'} \int d\tau \int_0^{2\pi \alpha' p^+} d\sigma \frac{1}{2}
(\dot{\vec{r}}^2-\vec{r}'^{2}-\vec{r}^2)
,}
with $\alpha' p^+=\frac{J}{g \sqrt{N}}$.
This part has the standard spectrum of four massive decoupled oscillators. On the other hand,
$\SS_{z^a,0}$ (for $a=1,2$) are each related to two of the four transverse coordinates that descend from the
$S^5$ part of the full geometry. We have
\eqn\sphizero{
\SS_{z^a,0}=\frac{1}{2\pi \alpha'} \int d\tau \int_0^{2\pi \alpha' p^+} d\sigma \frac{1}{2}
(\dot{z^a} \dot{\bar{z}^a} - {z'^a} {\bar{z}'^a}  + i (\bar{z}^a \dot{z^a} - z^a \dot{\bar{z}^a}))
.}
These parts give the Landau spectra that were discussed in section 2.

From the spectra that were derived in section 3, it is immediately clear how the above worldsheet actions should be
deformed to reproduce them. Since the operators with insertions of $D_i
\Omega$ and no phases are still protected in the deformed theory, being descendants
of $\Pi_J$, they continue to have $\Delta-J=1$. Hence, the corresponding part
of the worldsheet action involving the $\vec{r}$-part
still consists of four decoupled oscillators and it can
be written as
\eqn\srifull{
\SS_{\vec{r},h_1h_2}=\frac{1}{2\pi \alpha'} \int d\tau \int_0^{2\pi \alpha' p^+} d\sigma \frac{1}{2}
(\dot{\vec{r}}^2-\vec{r}'^{2}-\vec{r}^2)
.}
Nevertheless, consistency with the results of Table 1, requires the modified lightcone momentum
$p^+=\frac{J}{\alpha'g\sqrt{\FF_2}\sqrt{N}}$.

For the $z^a$-part we should reproduce the spectra associated to the anomalous
dimensions of eqs.\ \fullgammasolutionone, \fullgammasolutiontwo. These spectra together with the corresponding
gauge theory operators are summarized in Table 1. It is straightforward to
verify that they can be reproduced by the following worldsheet action
\eqn\phibosonicaction{\eqalign{
\sum_{a=1}^2 \SS_{z^a,h_1,h_2} &=\frac{1}{2\pi \alpha'} \sum_{a=1}^2
\int d\tau \int_0^{2\pi \alpha' p^+} d\sigma \frac{1}{2}
(\dot{z^a} \dot{\bar{z}^a} - {z'^a} {\bar{z}'^a}  +
\cr
& + i (\bar{z}^a \dot{z^a} - z^a \dot{\bar{z}^a})+
\alpha_1^2 z^a \bar{z}^a + \frac{1}{2}
(-1)^{a+1} \alpha_2 i ( z^a \bar{z}'^a -\bar{z}^a z'^a)
)
.}}

\vskip 5pt
\begintable
{\bf gauge theory operator} | ${\rm
{\bf \Delta-J}}$ \crthick $ \sum_l e^{il \varphi} \Omega ... \Omega \Psi^1
\Omega ... \Omega$ | $-1+\sqrt{1+\alpha_1^2+\alpha_2 \frac{g \sqrt{\FF_2}\sqrt{N}
n}{J}+ \frac{\FF_2 g^2 N n^2}{J^2}}$ \cr $\sum_l e^{il \varphi} \Omega ...
\Omega \bar{\Psi}^1 \Omega ... \Omega$ | $1+\sqrt{1+\alpha_1^2-\alpha_2
\frac{g \sqrt{\FF_2}\sqrt{N} n}{J}+ \frac{\FF_2 g^2 N n^2}{J^2}}$ \cr $\sum_l e^{il \varphi}
\Omega ... \Omega \Psi^2 \Omega ... \Omega$ |
$-1+\sqrt{1+\alpha_1^2-\alpha_2 \frac{g \sqrt{\FF_2}\sqrt{N} n}{J}+ \frac{\FF_2 g^2 N
n^2}{J^2}}$ \cr $\sum_l e^{il \varphi} \Omega ... \Omega \bar{\Psi}^2
\Omega ... \Omega$ | $ 1+\sqrt{1+\alpha_1^2+\alpha_2 \frac{g \sqrt{\FF_2}\sqrt{N}
n}{J}+ \frac{\FF_2 g^2 N n^2}{J^2}}$
\endtable

\noindent{\bf Table 1:}
{\sl
Anomalous dimensions associated to $\varphi$-dependent
insertions of  $\Psi^1, \bPsi^1, \Psi^2$
and $\bPsi^2$ in $\Pi_J$.
The fields appearing in the above operators are
the lowest bosonic components of the corresponding superfields of the previous
section.}

\vskip 4pt

Because of the way the minus sign of the $\frac{g \sqrt{\FF_2}\sqrt{N} n}{J}$ term
appears in the anomalous dimensions,
there is
a subtle difference in the correspondence between string states and
gauge theory
operators as it appears here and in eqs.\ \exampleczero\ and \bzerovac\ at the $\NN=4$ point. If we name $c^a$ and
$\bar{c}^a$ the oscillators of the worldsheet fields $z^a$ and $\bar{z}^a$ that appear in \phibosonicaction\
then with respect to the oscillators $b^a$ and $\bar{b}^a$ that appear in \exampleczero\ and \bzerovac\ we have
the twisted relation
\eqn\twistedcorrespondence{\eqalign{
& b^1 = c^2, \ \ \ \ b^2 = c^1,
\cr
& \bar{b}^1 = \bar{c}^1, \ \ \ \ \bar{b}^2 = \bar{c}^2.
}}

The bosonic worldsheet action of this section shows that the deformation has
turned on a 2-form NS-NS B-field with constant field strength
along the $(z^1,z^2)$ plane
and
has modified the metric accordingly. The validity of the supergravity equations of motion in addition requires the modification
of the 5-form field strength and/or the presence of a 3-form R-R flux. For the full determination of these components
of the dual background we also need to analyze the fermionic string sector, which we now proceed to do explicitly.
In this analysis it is convenient to ignore the supersymmetric partners of the ``magnetic'' terms of the bosonic
action \phibosonicaction. This means that implicitly we choose to work on the rotated coordinate system
of eq.\ \xplusrotation\ and the spectra we would like to reproduce in string theory do not involve the $\pm 1$ twist
outside the square root.

\subsec{Fermionic sector}

In this subsection we provide an analysis of the fermionic string sector. Our conventions follow closely
those of \myers\ and \mettseyt.
In the light-cone gauge, the fermionic part of the worldsheet action in the presence
of 5-form R-R and 3-form NS-NS and R-R fluxes becomes
\eqn\fermionicactiongeneral{\eqalign{
S_F=&-\frac{i}{\pi} \alpha' p^+ \int d\tau \int _0 ^{2\pi \alpha' p^+} \bigg (
\bar{\theta} \Gamma_{-} (\partial_{\tau} \theta + \rho \partial_{\sigma} \theta ) +
\frac{1}{8} \bar{\theta} \Gamma_{-} \not{\!\!H}_3 \rho \theta +
\cr
& \bar{\theta} \Gamma_{-} \not{\!\!F}_3 \rho_1 \theta + \frac{1}{240}
\bar{\theta} \Gamma_- \not{\!\!F}_5 \rho_0 \theta \bigg )
.}}
$\theta^I$ (for $I=1,2$) denote two 16-component Majorana-Weyl spinors.
$\rho$, $\rho_0$ and $\rho_1$ are two-dimensional gamma matrices, which can be expressed in terms
of the Pauli matrices $\sigma_i$ as
\eqn\rhomatrices{
\rho_0=i\sigma_2, \ \ \ \rho_1=\sigma_1, \ \ \ \rho=\rho_0 \rho_1 =\sigma_3.
}
We have also defined
\eqn\slashes{
\not{\!\!H}_3=H_{+ij}\Gamma_{ij}, \ \ \ \not{\!\!F}_3=F_{+ij}\Gamma_{ij}, \ \ \
\not{\!\!F}_5=F_{+ijkl}\Gamma_{ijkl}
.}
With latin characters $i,j,...$ we symbolize the transverse coordinates $\vec{r}$ $(i=1,2,3,4)$,
$z^a$ $(i=5,7)$ and $\bar{z}^a$ $(i=6,8)$.

From the form of the light-cone energies listed in Table 1, we
anticipate NS-NS and R-R 3-form fluxes with non-zero components only along
$+56$ and $+78$ and hence we set
\eqn\fieldsubstitutions{
H_{+56}=H_{56}, \ \ \ H_{+78}=H_{78}, \ \ \ F_{+56}=F_{56}, \ \ \ F_{+78}=F_{78}
}
and
\eqn\fiveformsubsti{
F_{+1234}=F_{+5678}=M
.}
Accordingly, the fermionic action \fermionicactiongeneral\ becomes
\eqn\sferm{\eqalign{
S_F=&-\frac{i}{\pi} \alpha' p^+ \int d\tau \int _0 ^{2\pi \alpha' p^+} \bigg (
\theta^1 \Gamma_{-} \partial_+ \theta^1 + \theta^2 \Gamma_{-} \partial_{-} \theta^2+
\cr
&+\frac{1}{2} \theta^1 \Gamma_- (F_{56}\Gamma_{56}+F_{78}\Gamma_{78}) \theta^2+
\frac{1}{4} \theta^1 \Gamma_- (H_{56}\Gamma_{56}+H_{78} \Gamma_{78}) \theta^1-
\cr
&- \frac{1}{4} \theta^2 \Gamma_- (H_{56} \Gamma_{56}+H_{78} \Gamma_{78}) \theta^2
-2M \theta^1 \Gamma_- \Gamma_{5678} \theta^2 \bigg )
}}
and $\partial_{\pm}=\partial_{\tau}\pm \partial_{\sigma}$.
By Fourier expanding the fermionic coordinates
\eqn\fourierfermionic{
\theta^I(\tau,\sigma)=\sum_n \theta_n ^I (\tau) e^{i \frac{n}{\alpha' p^+}\sigma}
}
and substituting into the equations of motion that derive from \sferm\ we get
\eqn\fermeom{\eqalign{
&\dot{\theta}^1_n + \frac{1}{4} \bigg(
F_{56}\Gamma_{56}+F_{78}\Gamma_{78} - 4 M \Gamma_{5678} \bigg) \theta^2_n +
\frac{1}{4} \bigg (H_{56}\Gamma_{56}+H_{78}\Gamma_{78} +4 i \frac{n}{\alpha' p^+} \bigg) \theta^1_n=0,
\cr
&\dot{\theta}^2_n + \frac{1}{4} \bigg(
F_{56}\Gamma_{56}+F_{78}\Gamma_{78} + 4 M \Gamma_{5678} \bigg) \theta^1_n -
\frac{1}{4} \bigg (H_{56}\Gamma_{56}+H_{78}\Gamma_{78} +4 i \frac{n}{\alpha' p^+} \bigg) \theta^2_n=0.
}}
Differentiating these equations with respect to $\tau$ and using them again to eliminate first derivatives
gives
\eqn\complexspinoreq{
\ddot{\varepsilon}_n + m_n \varepsilon_n =0,}
where
\eqn\epsilonmass{\eqalign{
&m_n =\frac{1}{16} \bigg \{ F_{56}^2+F_{78}^2 - 2F_{56}F_{78} \Gamma_{5678} + 16 M^2 +
\cr
& +  H_{56}^2+H_{78}^2 - 2 H_{56}H_{78} \Gamma_{5678}
- 8 i \frac{n}{\alpha' p^+} (H_{56}\Gamma_{56}+H_{78}\Gamma_{78} ) +
\frac{16n^2}{(\alpha' p^+)^2}  \bigg \}
.}}
We are following the notation of \myers\ and we have combined the spinors $\theta^1$ and $\theta^2$ into a single complex
spinor $\varepsilon=\theta^1+i \theta^2$.
Considering constant spinors $\varepsilon^{\pm \pm}$ with eigenvalues
\eqn\chevalier{\eqalign{
&i\Gamma_{56} \varepsilon ^{\pm (\cdot)} = \pm \varepsilon ^{\pm (\cdot)},
\cr
&i\Gamma_{78} \varepsilon^{(\cdot)\pm} = \pm \varepsilon ^{(\cdot)\pm}
}}
under the rotation generators $i \Gamma_{56}$ and $i \Gamma_{78}$ gives
\eqn\cheveigen{\eqalign{
&m_n \varepsilon^{++} = \frac{1}{16} \bigg \{ (F_{56}+F_{78})^2 + 16 M^2+
(H_{56}+H_{78})^2 - 8 \frac{n}{\alpha' p^+} (H_{56}+H_{78}) + \frac{16 n^2}{(\alpha' p^+)^2}
\bigg \} \varepsilon^{++},
\cr
&m_n \varepsilon^{+-} = \frac{1}{16} \bigg \{ (F_{56}-F_{78})^2 + 16M^2 +
(H_{56}-H_{78})^2 - 8 \frac{n}{\alpha' p^+} (H_{56}-H_{78}) + \frac{16 n^2}{(\alpha' p^+)^2}
\bigg \} \varepsilon^{+-},
\cr
&m_n \varepsilon^{-+} = \frac{1}{16} \bigg \{ (F_{56}-F_{78})^2 + 16 M^2+
(H_{56}-H_{78})^2 + 8 \frac{n}{\alpha' p^+} (H_{56}-H_{78}) + \frac{16 n^2}{(\alpha' p^+)^2}
\bigg \} \varepsilon^{-+},
\cr
&m_n \varepsilon^{--} = \frac{1}{16} \bigg \{ (F_{56}+F_{78})^2 + 16 M^2+
(H_{56}+H_{78})^2 + 8 \frac{n}{\alpha' p^+} (H_{56}+H_{78}) + \frac{16 n^2}{(\alpha' p^+)^2}
\bigg \} \varepsilon^{--}
.}}

We know already from the analysis of the bosonic sector that
\eqn\bfieldstrength{
H_{56}=-H_{78}=\alpha_2
.}
Hence, at each level $n$, we find four fermionic oscillators with equal frequencies
\eqn\rpartfrequency{
\omega_n = \sqrt{M^2 +\frac{1}{16}(F_{56}+F_{78})^2 + \frac{n^2}{(\alpha' p^+)^2} }
,}
which should be associated with the $r^i$-part of the background and the following
two pairs of frequencies
\foot{Remember that in each of these pairs there should be a $\pm 1$ twist, as required by the gauge theory values of
$\Delta-J$, but as noted in the previous subsection, we work here with the rotated coordinate system \xplusrotation,
where such twists are not supposed to appear.}
\eqn\phipartfrequency{\eqalign{
&\omega_n = \sqrt{M^2+\frac{1}{4}\alpha_2^2 +\frac{1}{16} (F_{56}-F_{78})^2
- \alpha_2 \frac{n}{\alpha' p^+} +
\frac{n^2}{(\alpha' p^+)^2}},
\cr
&\omega_n = \sqrt{M^2+\frac{1}{4} \alpha_2^2 +\frac{1}{16} (F_{56}-F_{78})^2
+ \alpha_2 \frac{n}{\alpha' p^+} +
\frac{n^2}{(\alpha' p^+)^2}}
,}}
which should be associated with the corresponding $(z^a,\bar{z}^a)$
directions of the dual spacetime.

Due to the fact that the background preserves some supersymmetry (as we
will verify explicitly in a moment), the above frequencies are expected to
be equal to the corresponding bosonic ones.
A direct comparison with the results expected from the
gauge theory side gives the following two equations for
the field strengths of the 3- and 5-form fluxes
\eqn\fluxconstraints{\eqalign{
&1 = M^2 +\frac{1}{16}(F_{56}+F_{78})^2,
\cr
&1+ \alpha_1^2 = M^2+\frac{1}{4}\alpha_2^2 +\frac{1}{16} (F_{56}-F_{78})^2
.}}
These constraints also imply the supergravity equations of motion of the next
subsection
and are not enough to fully determine the unknown constants $F_{56}$, $F_{78}$ and $M$.
This leaves an abundance of dual backgrounds with the required spectra\foot{Unless
we demand maximal supersymmetry the background is not unique
even at the $\NN=4$ point, where $a_1=a_2=0$.
}.
As we find in the next subsection,
all these backgrounds have the minimal sixteen supersymmetries of a pp-wave, except for a particular one that
has eight supernumerary supercharges.

\subsec{Spacetime geometry and supersymmetries}

The above discussion has led to a type IIB supergravity background with a non-zero R-R four-form potential $C_4$ and
self-dual field strength and non-zero NS-NS and R-R two-form potentials $B_2$ and $C_2$ respectively.
The analysis below follows closely the conventions of \refs{ \myers, \schwarz}.
As usual, we combine the two-form potentials
into a complex potential $A_2=B_2+iC_2$ with field strength $G_3=dA_2=H_3+iF_3$ and the self-dual
five-form is given by
\eqn\fiveformsugra{
F_5=\star F_5 = dC_4 -\frac{1}{8} {\rm Im} (A_2 \wedge G_3^*),
}
where $^*$ denotes complex conjugation and $\star$ ten-dimensional Hodge duality. The equations of motion
are
\eqn\sugraeoms{\eqalign{
R_{ab}=\frac{1}{6} F_{acdef}&F_b^{cdef} + \frac{1}{8} \bigg ( G_{acd}G_b^{*cd}+
G^*_{acd}G_b^{cd} - \frac{1}{6} g_{ab} G_{cde} G^{*cde} \bigg ) ,
\cr
d \star& G_3 = 4i F_5 \wedge  G_3 ,  \ \ \ \ G_{abc} G^{abc} =0,
\cr
&d \star F_5 = dF_5=-\frac{1}{8} {\rm Im} (G_3 \wedge G_3^*),
}}
together with the Bianchi identity $dG_3=0$.

In our case, the light-cone worldsheet theory \srifull\ and \phibosonicaction\ leads (after
a coordinate rotation of the type \xplusrotation) to a spacetime metric of the form
\eqn\fullmetric{
ds^2 = -4dx^+ dx^- - H(x) ({dx^+})^2 + \sum_{i=1}^8 dx^i dx^i
,}
with
\eqn\Hfunction{
H(x)= \sum_{i=1}^4 x^i x^i +(1+ \alpha_1^2) \sum_{i=5}^8 x^i x^i
.}
We also get a 5-form flux of the form
\eqn\fullfiveform{
F_5=(1+\star)dx^+ \wedge \omega_4
,}
with
\eqn\omegafour{
\omega_4 = M dx^1 \wedge dx^2 \wedge dx^3 \wedge dx^4
}
and a 3-form flux of the form $G_3=dx^+ \wedge \xi_2$ with
\eqn\xitwo{
\xi_2=\bigg (\alpha_2+iF_{56} \bigg ) dx^5 \wedge dx^6 +
\bigg ( -\alpha_2 +iF_{78} \bigg ) dx^7 \wedge dx^8
.}

The supergravity equations of motion \sugraeoms\ give the equation of motion
\eqn\sugraeomspecific{
\nabla^2 H = \frac{2}{3} \omega_4^2 + \frac{1}{2} |\xi_2|^2
,}
where $\nabla^2$ is the Laplacian in the transverse eight directions,
$\omega_4^2 = \omega_{ijkl}\omega^{ijkl}$ and $|\xi_2|^2=\xi_{ij}\xi^{*ij}$.
Using equations \Hfunction, \omegafour\ and \xitwo\ we get
\eqn\lastconstraint{
8(2+\alpha_1^2)=16M^2+2\alpha_2^2+F_{56}^2+F_{78}^2
.}
This equation is a consequence of \fluxconstraints\ and does not imply a new constraint on the 3- and 5-form
components. For the special value $F_{56}=-F_{78}=f$ we further get
\eqn\finalback{\eqalign{
&f^2=4\alpha_1^2-\alpha_2^2,
\cr
&M^2=1
.}}
In order for this ansatz to make sense, $4\alpha_1^2$ should be greater than $\alpha_2^2$. At first order
in the deforming parameters, we can
explicilty check that this inequality is always satisfied.

Now we would like to find the amount of supersymmetry preserved by the
above backgrounds. Our analysis follows closely \refs{\onepapado,\myers,
\fourpapado}. Since
the background is purely bosonic, the
supersymmetry variations of the dilatino and gravitino
take the form
\eqn\dilagravivara{\eqalign{
& \delta \lambda  = \frac{1}{24} G_{a b c} \Gamma^{a b c} \epsilon \cr
& \delta \psi_a  =  \DD_a \epsilon - \Omega_a \epsilon -
\Lambda_a \epsilon^*
}}
where\foot{$\Omega_a$ should not be confused with the superfield
$\Omega$ we used earlier.}
\eqn\omegalambda{\eqalign{
& \Omega_a = - \frac{i}{480} F_{bcdef} \Gamma^{bcdef} \Gamma_a \cr
& \Lambda_a =\frac{1}{96} \Big(G_{bcd}\Gamma_{a}^{bcd} -
9 G_{abc}\Gamma^{abc} \Big) ,
}}
and $\DD_a = \partial_a +  \frac{1}{4} \omega_{a \widehat{b}\widehat{c}}
\Gamma^{\widehat{b}\widehat{c}}$. Note that when necessary, we
distinguish tangent space indices from space-time ones by putting a hat on the former.

In terms of the non-coordinate basis
\eqn\newbasis{
e^{\tp} = dx^+, \;\;\; e^{\tm} = dx^- + \frac{1}{4} H(x) dx^+, \;\;\;
e^i=dx^i, i=1,\ldots,8
}
for the metric \fullmetric,
the only non-vanishing component of the spin
connection reads
\eqn\spin{
\omega_{\tp i} =  \frac{1}{2} \partial_i H(x) dx^+ .
}

For the background under consideration, the dilatino variation
is given by
\eqn\dilatinoone{
\delta\lambda = \frac{1}{8} \not{\!\!\xi_2 } \Gamma^{\tp} \epsilon ,
}
where $\not{\!\!\xi_2 } = (\xi_2)_{ij} \Gamma^{ij}$
and it can be written more explicitly
as
\eqn\dilatinotwo{
(1-\Gamma^0 \Gamma^9)\Big(\alpha_2 (\Gamma^{56}-\Gamma^{78})
+ i (F_{56}\Gamma^{56}+F_{78}\Gamma^{78})\Big) \epsilon = 0 .
}
We use the standard relations
$ \Gamma^{\widehat{\pm}} = \frac{1}{2}(\Gamma^0\pm
\Gamma^9)$
and $(\Gamma^i)^2 = (\Gamma^9)^2 = - (\Gamma^0)^2 = 1$.

It is possible to classify the solutions of \dilatinotwo\ by the eigenvalues
of the mutually commuting Lorentz generators
$\Gamma^0\Gamma^9, i \Gamma^1\Gamma^2,
i\Gamma^3\Gamma^4, i\Gamma^5\Gamma^6, i\Gamma^7\Gamma^8$. Notice
that the dilatino variation is independent of the eigenvalues
of $ i \Gamma^1\Gamma^2, i\Gamma^3\Gamma^4$ but because of the chirality
constraint $\Gamma^{11} \epsilon =\epsilon$
we eventually have only a two-fold degeneracy in the
number of (complex) solutions.

The standard 16 supersymmetries preserved by generic pp-waves
correspond to the spinors with $\Gamma^0\Gamma^9 = 1$, i.e.
they are annihilated by  $\Gamma^{\tp}$.
We show explicitly
in a moment that the gravitino variation is also zero for
these. For 3-form R-R fluxes
satisfying $F_{56} + F_{78} \neq 0$ we can't obtain any more supersymmetries.
For the special case  $F_{56}=-F_{78}=f$, however, we get 8 supernumenary
Killing spinors
with $ i\Gamma^5\Gamma^6 = i\Gamma^7\Gamma^8 = \pm$. Hence, this particular string
background preserves 24 supersymmetries.

In order to show that the extra supersymmetries are indeed preserved,
we also have to verify that the corresponding gravitino variation vanishes
\eqn\gravitino{
\DD_a \epsilon = \Omega_a \epsilon + \Lambda_a \epsilon^* .
}
For the general background
we have\foot{Notice that in our conventions the non-zero components of the
tangent space metric read
$\eta_{\tp \tm} = -2, \; \eta^{\tp \tm}=-1/2$ and accordingly
$\Gamma^{\tp} \Gamma^{\tm} + \Gamma^{\tm} \Gamma^{\tp} = -1, \;\;
\Gamma_{\tp} \Gamma_{\tm} + \Gamma_{\tm} \Gamma_{\tp} = -4 .$
In addition, $\Gamma_{+} = \Gamma_{\tp} + \frac{1}{4} H(x) \Gamma_{\tm}, \;\;
\Gamma_{-} = \Gamma_{\tm}, \;\;\Gamma^{+} = \Gamma^{\tp},\;\;
\Gamma^{-} = \Gamma^{\tp} - \frac{1}{4} H(x) \Gamma^{\tp}$. We
frequently use these relations in the ensuing.}
\eqn\omegalambdaour{\eqalign{
& \Omega_a = -i \frac{M}{4} (\Gamma^{1234}+\Gamma^{5678}) \Gamma^{\tp}
\Gamma_a \cr
& \Lambda_a = \frac{1}{32}\Big((\xi_2)_{ij} \Gamma_{a}^{\;\tp ij} -
3 G_{abc} \Gamma^{bc} \Big) .
}}

Since $\Gamma^{\tp} \Gamma_{\tm} = 0$ and $G_{- ab} = 0$,
it is easy to see that $\Omega_- = \Lambda_- = 0$. In addition, the component
of the spin connection along $dx^-$ is zero. Hence, the Killing spinors
are independent of $x^-$ and we can write $\epsilon=\epsilon(x^+,x^i)$.

For the $i$ components, on the other hand, we get
\eqn\icomponents{\eqalign{
& \Omega_i = -i \frac{M}{4} (\Gamma^{1234}+\Gamma^{5678}) \Gamma^{\tp}
\Gamma_i \cr
& \Lambda_i = \frac{1}{32}\Big( \Gamma_i  \not{\!\!\xi_2 }
- 8 (\xi_2)_{ij}  \Gamma_{j}\Big) \Gamma^{\tp} .
}}
Using the identity $(\Gamma^{\tp})^2 = 0$ we find that $\Omega_i \Omega_j =
\Lambda_i \Lambda_j = \Omega_i \Lambda_j = 0$ and
from $\partial_i \epsilon = \Omega_i \epsilon
+\Lambda_i \epsilon$ we conclude that $\Omega_i \epsilon$
and $\Lambda_i \epsilon^*$ are $x^i$-independent and, accordingly, that
the Killing spinors take the form
$\epsilon = \epsilon_{1}(x^+) + x^i \epsilon_{2}(x^+)$. From this we
further get $\partial_i \epsilon = \epsilon_2$ and we can eventually set
\eqn\kilspin{
\epsilon = \epsilon_1 + x^i (\Omega_i \epsilon_1 + \Lambda_i \epsilon_1^*) .
}
Note that the dilatino variation now takes
the form $\not{\!\!\xi_2 } \Gamma^{\tp} \epsilon_1 =0$.

For the $a=+$ component of \gravitino\ we have
\eqn\gravitinoplus{
\partial_+ \epsilon = \frac{1}{4} \partial_i H(x) \Gamma^i
\Gamma^{\tp} \epsilon + \Omega_{+} \epsilon + \Lambda_{+} \epsilon^*
}
where
\eqn\omegalambdaplus{\eqalign{
& \Omega_{+} = -i \frac{M}{4} (\Gamma^{1234}+\Gamma^{5678}) \Gamma^{\tp}
\Gamma_{\tp} \cr
& \Lambda_{+} = -\frac{1}{32} \not{\!\!\xi_2 } (\frac{1}{2}
 \Gamma_{\tp}  \Gamma_{\tm} + 4) .
}}

Our background has $H(x) = H_{ij} x^i x^j$ with
$H_{ij} = \delta_{ij}, {\rm when} \;i,j=1,\ldots,4, \;\;
H_{ij} = (1+a_1^2) \delta_{ij}, {\rm when} \;
i,j=5,\ldots,8$ and zero otherwise.
Plugging-in the specific form of the Killing spinors \kilspin\
into \gravitinoplus\ and collecting the terms independent of the $x^i$
gives
\eqn\xiindependent{
\partial_+ \epsilon_1 =  -i \frac{M}{2} (\Gamma^{1234}+\Gamma^{5678})
\epsilon_1  -\frac{1}{32} \not{\!\!\xi_2 } (\frac{1}{2}
 \Gamma_{\tp}  \Gamma_{\tm} + 4) \epsilon_1^* .
}
We used the equation
$ (\Gamma^{1234}+\Gamma^{5678})  \Gamma^{\tm} \Gamma^{\tp} \epsilon_1 = 0$,
which follows from the chirality constraint.
Substituting this result into the equation we get from the $x^i$-linear terms of \gravitinoplus\
gives
\eqn\algebraic{\eqalign{
&
\Big( -H_{ij} \Gamma^j \Gamma^{\tp}  + M^2 \Gamma^i \Gamma^{\tp} + \frac{1}{4}
 \slash{\xi_2} \Lambda_i^* - \frac{1}{8} \Lambda_i  \slash{\xi_2}^*
\Big)\epsilon_1 + \cr
&
\Big( +i M \{(\Gamma^{1234}+\Gamma^{5678}),\Lambda_i\}-\frac{1}{8}
\Omega_i \slash{\xi_2} -\frac{1}{4}  \slash{\xi_2} \Omega_i
\Big)\epsilon_1^* = 0 .
}}
After some $\Gamma$-matrix technology we further obtain
\eqn\algebraicnew{
\Big( -H_{ij} \Gamma^j  + M^2 \Gamma^i + \frac{1}{32} \xi_{jk} \xi^*_{kl}
\Gamma_i \Gamma_{jl} + \frac{1}{4} \xi^*_{ij} \xi_{jk}\Gamma_k\Big)
 \Gamma^{\tp} \epsilon_1 -
 \frac{i M}{8} \Gamma^{1234} \Gamma_i  \slash{\xi_2}
 \Gamma^{\tp} \epsilon_1^* = 0 .
}
To derive the above equation we make use of the identities
$(\Gamma^{1234}+\Gamma^{5678}) \Gamma^{\tp} \epsilon_1 = 0$ and
$\Gamma^{1234} \Gamma_i  \Gamma^{\tp}  \epsilon_1 =
+\Gamma^{5678} \Gamma_i  \Gamma^{\tp}  \epsilon_1$ that hold because of
the chirality condition and the relation $[\Gamma_i,  \slash{\xi_2}] =
4 \xi_{ij} \Gamma_j$.
In addition, we exploit the fact that our
background has non-zero $\xi_{ij}$ components only along the $x^5, x^6, x^7, x^8$
directions and hence
$[ \Gamma^{1234}, \slash{\xi_2}]
= [ \Gamma^{5678}, \slash{\xi_2}] = 0$.

It is clear from eq.\ \algebraicnew\ that the 16 supersymmetries annihilated
by $\Gamma^{\tp}$ are preserved, as expected for generic pp-waves.
Moreover, recall that
for $F_{56}=-F_{78}=f$ the dilatino variation gave eight extra potential
Killing spinors. For this ansatz $\slash{\xi_2} = 2
(a_2+i f) (\Gamma^{56} - \Gamma^{78})$ and the supernumenary
spinors satisfy $ (\Gamma^{56} - \Gamma^{78}) \epsilon_1 = 0$.
As a result of this,
$\slash{\xi_2}\Gamma^{\tp} \epsilon_1^* = 0$ and
the second term of \algebraicnew\ vanishes.
The first term of \algebraicnew\ is also zero.
For $i=1,2,3,4$ it vanishes
on the supernumenary spinors provided that $-1+M^2 =0$, whereas
for  $i=5,6,7,8$ it vanishes when
$-(1+a_1)^2 + M^2 +
\frac{1}{4}|a_2+i f|^2 = 0$. Both of these constraints hold, because
of the equations of motion  \finalback\ .

We conclude that the reverse-engineering process that is being employed in this paper
does not produce a unique ``dual'' background at the infinite $J$ limit, but curiously enough,
it produces a unique pp-wave string
background with the right spectrum and supersymmetry enhancement to 24 supersymmetries.
The same process gives a similar abundance
of ``dual'' backgrounds at the $\NN=4$ point as well, but only one of them has maximal supersymmetry and
results as the Penrose limit of the full $AdS_5 \times S^5$ geometry.

\newsec{Conclusions and future directions}

In this paper we considered the extension of the BMN correspondence for $\NN=1$ superconformal Yang-Mills
theories that are obtained as exactly marginal deformations of the $\NN=4$ theory. We concentrated on the gauge
theory side of this correspondence and analysed the effect of the deformation on the large R-charge BMN operators
of the $\NN=4$ point. First, we noticed that the deforming superpotential breaks the $SO(6)$ R-symmetry group into
a $U(1)$, under which all three of the complex Higgs fields are equally charged. For that reason, it was more natural to
express the $\NN=4$ BMN correspondence in terms of a ``magnetic'' pp-wave. With this $U(1)$ we reconsidered the
BMN operators both before and after the exactly marginal deformations of the $\NN=4$ theory.
On the assumption that such operators are quasi-primary, we used the $\NN=1$
superspace techniques of \italian\ to derive their anomalous dimensions for finite deforming parameters
and verified this result in leading order in perturbation theory.

The picture we find is the following. At the $\NN=4$ point we have an infinity of degenerate operators at each level,
which are mapped on the string theory side to a corresponding infinite set of Landau-degenerate states. The effect
of the deformation on the gauge theory side is to break the $\NN = 4$
short multiplets into $\NN=1$ short and long ones and give
anomalous dimensions to many previously protected operators. As a result, the $\NN=4$ Landau degeneracy is lifted
and we are only left with the three chiral protected operators $\tr [\Omega^J], \tr [{(\Psi^1)}^J]$ and
$\tr [{(\Psi^2)}^J]$. At the $\NN=4$ point we presented the BMN correspondence by using appropriate insertions into the
first of these operators only. This works in a natural way
at the $\NN=4$ point, but less obviously after the deformations, because
the Landau degeneracy has been lifted.
As for the other previously protected operators with either $\Delta-J=0$ or
$\Delta-J=2$,
we find that they acquire anomalous dimensions
which are functions of the deforming parameters.

After the determination of the above anomalous dimensions we ask how we can reproduce them as spectra of an
appropriately defined light-cone worldsheet theory. The answer to this question turns out not to be unique, but in all cases
we find that the spacetime effect of the deformation is to modify the trace of the transverse metric and
turn on 3-form R-R and NS-NS fields. This seems to be consistent with the first and second order analysis
of these deformations in the full geometry \refs{\novel,\aharony}.
It would be interesting to work out explicitly the Penrose
limit of these backgrounds and compare with what was obtained
in section 4.

Concerning the non-uniqueness of the resulting backgrounds
we believe that this is an artifact of the reverse-engineering process at the infinite $J$ limit. It is natural to expect
that the extension of our analysis at finite $J$ will produce a unique dual background. It would be interesting to see if
this background is the finite $J$ version of the unique pp-wave with twenty-four supernumerary supersymmetries that
we find from the infinite $J$ reverse-engineering process.

We consider the analysis of this paper as a first step analysis of the BMN correspondence for
conformal $\NN=1$ SYM theories.
There is an interesting set of subjects that one could further explore. First, it would be nice to understand better how
the short and long $SU(2,2|4)$ representations break up into short and long $SU(2,2|1)$ representations as we turn on
the marginal deformations. It is clear from our analysis that at the $\NN=4$ point protected operators
belong to certain $\NN=1$ short multiplets, which break into long multiplets after the deformation and the previously
protected operators become unprotected. This deformation of the multiplets is also evident in the large $J$ version of the
equations of motion \deformeduoeom. A different effect is the following.
The $\NN=4$ analysis of \beisert\ focused on single trace BMN operators
with two insertions and dimension $\Delta$ and showed that they belond in $[\Delta/2]$
long multiplets, thus proving the equality of the anomalous dimensions of several BMN operators. A similar equality
is absent after the $\NN=4$-breaking deformations and one would like to understand better how these long
BMN multiplets rearrange themselves.

Another interesting direction would be
to extend our analysis at finite R-charge, as it was done for the $\NN=4$ theory in
\refs{\beisert,\andryzhov}. This will put the correspondence into firmer ground and will allow us to see if and how
the reverse-engineering process can produce a unique dual background. It would be nice if such a process, combined with
supersymmetry, could lead to the full dual supergravity background of the $\NN=1$ theories, which is expected to be a warped
fibration of $AdS_5$ over a deformed $S^5$ along with 3- and 5-form fluxes.
We expect, however, that such a process will be considerably complicated.

Finally, the author of \beisert\ made the interesting observation that the $J=0$ BMN supermultiplet at the $\NN=4$ point
coincides with the Konishi multiplet and that the large $J$ operators behave like generalized Konishi operators. He also
suggested that the BMN classification of operators based on the number of defects could be valid more generally and
might provide an alternative route towards a better understanding of
the full spectrum of the
$\NN=4$ theory and the related AdS/CFT correspondence.
It seems very possible that the same is true for the $\NN=1$ theories in the Leigh-Strassler moduli space and
it would be worthwhile to explore this possibility further.

\bigskip
\noindent{\bf Acknowledgements}

We would like to thank G. Dall'Agata, J. Erdmenger, Z. Guralnik,
A. Hanany, D. Kutasov, W. McElgin, A. Parnachev, A. C. Petkou, D. Sahakyan
and C. Sieg for useful discussions and D. Brecher, D.Z.  Freedman,
C. V. Johnson and  H. Osborn for
helpful correspondence.
The work of V. N. is supported by the DOE grant DE-FG02-90ER40560.
The work of N. P. is supported by the Deutsche Forschungsgemeinschaft under the
project number DFG Lu 419/7-2.


\listrefs
\bye